\documentclass[twoside,12pt]{article}

\usepackage{amssymb}
\usepackage{epsfig}
\usepackage{amsthm}
\usepackage{amsmath, mathrsfs,psfrag}

\newlength{\dinwidth}
\newlength{\dinmargin}
    
\newcommand{\C}{S} 
\newcommand{\DC}{\hat{S}} 
\newcommand{\Ab}{A}
\newcommand{\Spt}{\mathrm{Sp}}

\newcommand{\lmfa}{\mfa_{\te{loc}}}
\newcommand{\tracabd}{B(\hil)_{*,\te{bd}} }
\newcommand{\GG}{\mathfrak{g}}
\newcommand{\HH}{\mathfrak{h}}
\newcommand{\Ran}{\te{Ran}}

\newcommand{\Om}{\Omega}
\newcommand{\ccdot}{\,\cdot\,}
\newcommand{\tmfa}{\wt{\mfa}}
\newcommand{\thil}{\wt{\hil}}

\newcommand{\funq}{\underline{\fun}}
\newcommand{\alq}{\underline{\al}}

\newcommand{\D}{\mathcal{D}}

\newcommand{\we}{\te{w$^*$-}}
\newcommand{\no}{\mathrm{n}\te{-}}

\newcommand{\traca}{\trace^{(\te{a}) }}

\newcommand{\cl}{\mathrm{cl}}

\newcommand{\co}{\te{c}}
\newcommand{\dd}{d}

\newcommand{\f}{f}

\newcommand{\cc}{\scc}

\newcommand{\ovomi}{\om^{\mathbb{I}}}

\newcommand{\B}{L^{(1)}}
\newcommand{\CC}{\mfa^{(1)}}

\newcommand{\nin}{\noindent}
\newcommand{\hmfa}{\hat{\mfa}}
\newcommand{\te}{\textrm}
\newcommand{\PC}{\mathfrak{P}}

\newcommand{\Sp}{\mathrm{Sp}}

\newcommand{\sic}{\mathrm{sc}}

\newcommand{\un}{\underline}

\newcommand{\Span}{\textrm{Span}}
\newcommand{\Spant}{\textrm{\emph{Span}}}

\renewcommand{\d}{d}

\newcommand{\FH}{\mathrm{FH}}

\newcommand{\ac}{\mathrm{ac}}
\newcommand{\spp}{\mathrm{pp}}
\newcommand{\scc}{\mathrm{c}}

\newcommand{\wt}{\widetilde}

\newcommand{\tg}{\tilde{g}}

\newcommand{\h}{\half}

\newcommand{\tf}{\tilde{f}}

\newcommand{\om}{\omega}

\newcommand{\si}{\sigma}
\newcommand{\al}{\alpha}
\newcommand{\be}{\beta}

\newcommand{\la}{\lambda}
\newcommand{\vp}{\varphi}
\newcommand{\Ga}{\Gamma}
\newcommand{\ga}{\gamma}
\newcommand{\eps}{\varepsilon}
\newcommand{\De}{\Delta}

\newcommand{\nat}{\mathbb{N}}

\newcommand{\real}{\mathbb{R}}

\newcommand{\trace}{B(\hil)_*}  
\newcommand{\tracei}{B(\hil)_{*,\infty}}

\newcommand{\vep}{\vec{p}}

\newcommand{\veq}{\vec{q}}

\newcommand{\vx}{\vec{x}}

\newcommand{\ov}{\overline}

\newcommand{\fun}{\vp}

\newcommand{\su}{\substack}

\newcommand{\hil}{\mathcal{H}}
\newcommand{\mfa}{\mathfrak{A}}
\newcommand{\mco}{\mathcal{O}}

\newcommand{\cone}{\overline{V}_+}

\newcommand{\supp}{\textrm{supp}}
\newcommand{\fr}[2]{\frac{#1}{#2}}

\newcommand{\non}{\nonumber}

\newcommand{\vac}{\Omega}

\newcommand{\half}{\fr{1}{2}}
\newcommand{\lan}{\langle}
\newcommand{\ran}{\rangle}

\def\proof{\noindent{\bf Proof. }}
\def\qed{$\Box$\medskip}

\newtheorem{theoreme}{Theorem} [section]%
\newtheorem{proposition}[theoreme]{Proposition}%
\newtheorem{lemma}[theoreme]{Lemma}%
\newtheorem{definition}[theoreme]{Definition}%
\newtheorem{corollary}[theoreme]{Corollary}%
\newtheorem{remark}[theoreme]{Remark}%
\newtheorem{example}[theoreme]{Example}%
\newtheorem{criterion}[theoreme]{Criterion}%
\newcommand{\beq}{\begin{equation}}
\newcommand{\eeq}{\end{equation}}
\newcommand{\beqa}{\begin{eqnarray}}
\newcommand{\eeqa}{\end{eqnarray}}
\newcommand{\ben}{\begin{arabicenumerate}}
\newcommand{\een}{\end{arabicenumerate}}
\newcommand{\bex}{\begin{example}}
\newcommand{\eex}{\end{example}}
\newcommand{\ber}{\begin{remark}}
\newcommand{\eer}{\end{remark}}
\newcommand{\bec}{\begin{corollary}}
\newcommand{\eec}{\end{corollary}}
\newcommand{\bep}{\begin{proposition}}
\newcommand{\eep}{\end{proposition}}
\newcommand{\becr}{\begin{criterion}}
\newcommand{\eecr}{\end{criterion}}

\def\bel{\begin{lemma}}
\def\eel{\end{lemma}}
\def\bet{\begin{theoreme}}
\def\eet{\end{theoreme}}
\def\bed{\begin{definition}}
\def\eed{\end{definition}}

\begin{document}

\title{Continuous Spectrum of Automorphism Groups \\ and the Infraparticle Problem}

\author{ 
Wojciech Dybalski\\[5mm] 
  { Zentrum Mathematik,}
{ Technische Universit\"at M\"unchen,}\\ [2mm] 
{  D-85747 Garching, Germany}\\[2mm] 
e-mail: dybalski@ma.tum.de}

\date{}
\maketitle
\begin{abstract}

This paper presents a general framework for a refined 
spectral analysis of a group of isometries acting on a Banach space,
which extends the spectral theory of Arveson.
The concept of continuous Arveson spectrum is introduced and the corresponding spectral subspace is defined.   
The absolutely continuous and  singular-continuous 
parts of this spectrum are  specified. Conditions are given, in terms of the transposed action of the group of isometries, 
which guarantee that the pure-point and continuous subspaces span the entire Banach space. 
In the case of a unitarily implemented group of automorphisms, acting on a $C^*$-algebra,
relations  between the continuous spectrum of the automorphisms and the spectrum of the implementing
group of unitaries are found. The group of spacetime translation automorphisms in quantum field theory is analyzed
in detail. In particular, it is shown that the structure of its continuous spectrum is relevant to the problem
of existence of (infra-)particles in a given theory.

\end{abstract}
\section{Introduction}\label{chapter:introduction}
\setcounter{equation}{0}

In the familiar case of a strongly continuous group of unitaries $U(t)=e^{iHt}$, acting on a Hilbert space $\hil$, spectral theory is  well understood.
In particular, $\hil$ can be decomposed into the pure-point,  
absolutely continuous and singular-continuous subspaces, which  reflects the decomposition
of the spectral measure of $H$ into measure classes. 
On the other hand, for a strongly continuous group of isometries $\al_t=e^{iDt}$ acting
on  a Banach space $\mfa$,  spectral theory is much less developed. 
While the Arveson spectral theory  provides subspaces associated  with closed subsets of the spectrum of $D$ \cite{Ar82, Ev76,Lo77}, 
there does not seem to exist any general definition of the continuous spectral subspace, not to speak of its 
absolutely continuous or singular continuous parts. It is the main goal of the present paper to introduce such  notions and demonstrate 
their  relevance to the problem of particle interpretation  in quantum field theory (QFT).

In the Hilbert space setting these detailed spectral concepts    
provided a natural framework for the formulation and resolution of the problem of asymptotic completeness in quantum mechanics 
\cite{En78,SiSo87,Gr90,De93}. 
The absence of such structures on the side of Banach spaces impedes  the study 
of particle aspects in quantum field theory, where the time evolution is governed by a group of automorphisms $\al$ acting on
a $C^*$-algebra of observables $\mfa$. Asymptotic completeness is an open problem in all known models of interacting quantum fields, except
for a recently constructed class of two-dimensional theories with factorizing $S$-matrices \cite{Le08}. Since pairs of
charged particles may be produced in collisions of neutral particles, it is  not even \emph{a priori} clear what particle types a
given theory describes. Finally, the possible presence of charges with weak localization properties, like the electric charge in 
quantum electrodynamics, forces one to depart from the conventional Wigner concept of a particle as a state in some irreducible representation space of the Poincar\'e group  \cite{Wi39}. In spite of decades of research  \cite{Sch63, FMS79, Bu86},  this \emph{infraparticle problem} is still a largely open issue.

One approach to the problems listed above is to study simplified models which capture some relevant features
of quantum field theories. Over the last decade there has been significant progress along these lines
\cite{DG00,FGS04,Sp04,Pi05,CFP07,He07,Re09}. A complementary approach, pursued in algebraic quantum field theory, aims at a development
of a model-independent concept of a particle which is sufficiently general to encompass all the particle-like structures
appearing in quantum field theory \cite{Bu94}. Substantial steps in this direction were made by  Buchholz, Porrmann and Stein \cite{BPS91}.
In order to clarify the relation between the particle aspects of quantum field theory and the (Arveson) spectrum of the group of automorphisms $\real^{s+1}\ni (t, \vx)\to\al_{(t,\vx)}$, which describes spacetime translations of observables, we recall the main steps of this analysis: 
To extract the particle content of a physical state $\om\in\mfa^*$, one has to compensate for
dispersive effects. To this end,  one  paves the whole space with observables and sums up the results. This amounts to studying
the time evolution of the integrals 
\beq
\si_{\om}^{(t)}(A):=\int d^sx\, \om(\al_{(t,\vx)}(A)),  \label{asymptotic-functional-def}
\eeq
where $A\in\mfa$ is a suitable observable. The limit points of $\si_{\om}^{(t)}$ as $t\to\infty$, called the asymptotic 
functionals $\si_{\om}^{(+)}$,
carry information about all the particle types appearing in the theory.
In fact, for theories of Wigner particles it was shown in \cite{AH67}  that each 
asymptotic functional can be represented as a mixture of plane wave configurations of all the particle types 
contributing to the state $\om$
\beq
\si^{(+)}_{\om}(A)=\sum_{\la}\int d^sp\,\rho_{\la}(\vep)\,\lan \vep,\la |A| \vep,\la\ran.
\label{weltformel}
\eeq
This mixture is labeled by $\la=[m,s,q]$ (i.e. mass, spin and charge), possibly including pairs 
of charged particles. The functions $\rho_{\la}$  stand for the  asymptotic densities  of the respective particle types.  
In the general case, including both  Wigner particles and infraparticles,  
a similar expression was derived by Porrmann \cite{Po04.2}
\beqa
\si^{(+)}_{\om}(A)=\int d\mu(\la)\, \si^{(+)}_{\la}(A). \label{weltformel1}
\eeqa
Here the analogues of the plane wave configurations are the so-called pure particle weights $\si^{(+)}_{\la}$ labeled by $\la=[p,\gamma]$, 
where $p$ is the four-momentum of the particle and  $\ga$ carries information about the other
quantum numbers, like spin or charge. Each pure particle weight gives rise to an irreducible representation of $\mfa$. 
In contrast to the previous case, the representations corresponding to different four-momenta $p$ may  be inequivalent.
Thus the  infraparticle situation, e.g. the electron whose velocity gives a superselection rule 
\cite{FMS79,Bu82,Bu86}, can be treated in this framework.


A necessary prerequisite for this approach is the existence of non-zero asymptotic functionals. 
It is well known from the study of generalized free fields that this property does not follow from the general postulates of quantum field theory.
However, it has been established for theories of Wigner particles \cite{AH67} and in a non-interacting
model of an infraparticle introduced by Schroer \cite{Sch63,Jo91}. Moreover,  in \cite{Dy08.3} we supplied a model-independent argument, ensuring the existence of non-zero asymptotic functionals in a certain class of theories containing a stress-energy tensor\footnote{The relevance of the stress-energy tensor to particle aspects was first pointed out in \cite{Bu94}.}. In view of this recent progress, one can hope for a general classification of quantum field theories w.r.t. their particle structure. It is the main goal of the present work to develop a natural language for such a classification.

As a first step in this direction, we infer from formula~(\ref{weltformel}) that
the asymptotic functional is non-trivial, only if its domain contains sufficiently many observables, whose energy-momentum transfer 
includes zero.
We recall that the energy-momentum transfer of an observable $A$  coincides with the Arveson spectrum 
of the group of automorphisms $\al$ restricted to the subspace spanned by the orbit of $A$  (cf. definition~(\ref{Arveson-spectrum})). Therefore, 
essential information about the particle content of a given theory should be encoded in the properties of the Arveson spectrum of $\al$ 
in a neighborhood  of zero. As it is evident from definition~(\ref{asymptotic-functional-def}) that the joint eigenvectors of the generators of 
$\al$ do not belong to the domains of the asymptotic functionals, the relevant part of the spectrum is the continuous one.

However, the existing spectral theory is not yet sufficiently developed to test such fine features of the Arveson spectrum. It 
provides a functional calculus and spectral mapping theorems
for some classes of functions as well as a definition and properties of spectral subspaces associated with closed subsets of
the spectrum. (See \cite{Ar82} for a review). This allows e.g. for a study of the pure-point spectrum 
which has attracted much attention from various perspectives \cite{Ba78,Jo82,AB97,Hu99}. 
To our knowledge, there is no systematic analysis of the 
continuous Arveson spectrum in the literature. Nevertheless,  there exist  some interesting  results  pertaining to  decay properties
of the functions $\real^{d}\ni x\to\om(\al_x (A))$ and regularity properties of their Fourier transforms $\real^{d}\ni p\to\om(\wt{A}(p))$. We mention the analysis of Jorgensen \cite{Jo92}, inspired by the Stone formula, and  a result of Buchholz \cite{Bu90}, concerning
space translations in QFT, which will be used in Section~\ref{QFT-section} of the present paper.
The Fourier transforms $\om(\wt{A}(\,\cdot\,))$ appear also
as a tool in the  literature related to the Rieffel project of extending the notions of proper action and orbit space from the setting of group actions on locally compact spaces to the context of $C^*$-dynamical systems \cite{Ri90, Ex99, Ex00, Me01}. 
This recent revival of interest in  the subject is an additional motivation for the general analysis of the continuous Arveson spectrum which we undertake in this work.

Let $(\al, \mfa)$ be a strongly continuous group of isometries acting on a Banach space $\mfa$ and let $(\al^*,\mfa_*)$ be the transposed
action of $\al$ on a suitable closed, invariant subspace $\mfa_*\subset\mfa^*$. In this framework we define  the
pure-point and continuous spectral subspaces of $\al$ and $\al^*$, denoted by $\mfa_{\spp}$, $\mfa_{\cc}$ and $\mfa_{*,\spp}$, $\mfa_{*,\cc}$, respectively.
Certainly,  $\mfa_{\spp}$, (resp. $\mfa_{*,\spp}$), is spanned by the joint eigenvectors of the generators of $\al$, (resp. $\al^*$). 
In the absence of orthogonality, our definition of the continuous subspace is motivated by the Ergodic Theorem 
from the setting of groups of unitaries:   
\beq
\mfa_{\co}:=\{\, A\in\mfa\,|\, \underset{\su{q\in\real^\dd \\ \fun\in\mfa_*} }\forall \ 
\underset{K\nearrow \real^{\dd}}\lim\fr{1}{|K|}\int_{K} e^{-iqx}\fun(\al_x(A))\,d^{\dd}x=0\, \}. \label{continuous-subspace-intro}
\eeq
The subspace $\mfa_{*,\cc}$ is obtained by exchanging the roles of $\mfa$ and $\mfa_*$ above. 
In analogy with the Hilbert space setting, 
\beq
\lan \mfa_{*,\spp},\mfa_{\cc}\ran=\lan \mfa_{*,\cc},\mfa_{\spp}\ran=0,
\eeq
where $\lan\,\cdot\,,\,\cdot\,\ran$ denotes the evaluation of functionals from $\mfa^*$ on elements of $\mfa$. Exploiting these facts,
we find necessary and sufficient conditions for the following decompositions to hold
\beqa 
\mfa&=&\mfa_{\spp}\oplus\mfa_{\cc}, \label{point+cont}\\
\mfa_{*}&=&\mfa_{*,\spp}\oplus\mfa_{*,\cc}, \label{point*+cont*}
\eeqa
as well as examples, where these equalities fail.  
More refined analysis of the continuous spectrum requires the choice of some norm dense subspaces  $\hmfa\subset\mfa$ 
and $\hmfa_{*}\subset\mfa_{*}$. By definition, the absolutely continuous part $\mfa_{\ac}$ is generated by all  $A\in\mfa_{\cc}$
s.t. the Fourier transforms of the corresponding functions $\real^{\dd}\ni x\to\fun(\al_x(A))$ are  
integrable for all $\fun\in\hmfa_{*}$.  As for the singular-continuous space, in the Banach space setting the canonical choice 
is the quotient
\beq
\mfa_{\sic}:=\mfa_{\cc}/\mfa_{\ac}
\eeq
on which there acts the reduced group of isometries $\alq$, defined naturally on the equivalence classes. 

The point-spectrum  $\Sp_{\spp}\al$ consists of joint eigenvalues of the generators of $\al$. 
The continuous spectrum of $\al$, denoted by $\Sp_{\cc}\al$, is defined as the Arveson spectrum of $\al$
restricted to the subspace $\mfa_{\cc}$. The  absolutely continuous spectrum $\Sp_{\ac}\al$ is constructed
analogously. The singular-continuous part $\Sp_{\sic}\al$ is specified as the Arveson spectrum of $\alq$. The corresponding spectral concepts 
on the side of $\al^*$ are defined in obvious analogy.
We remark that in the Hilbert space setting the above spectra
and the subspaces $\mfa_{\spp}$, $\mfa_{\cc}$ and $\mfa_{\ac}$ coincide with the standard ones. The space $\mfa_{\sic}$ is 
isomorphic, as a Banach space, to the conventional singular-continuous subspace.

We are primarily interested in the case of a group of automorphisms $\al$ acting
on a $C^*$-algebra $\mfa$. We assume the existence of a pure state $\om_0$ on $\mfa$  s.t.
\beq
\ker\,\om_0\subset\mfa_{\cc}, \label{key-assumption-two}
\eeq 
if $\mfa_*$ is chosen as the predual of the GNS representation $(\pi,\hil,\Om)$ induced by $\om_0$. Then
there follow decompositions~(\ref{point+cont}) and (\ref{point*+cont*}) as well as the invariance
of $\om_0$ under the action of $\al^*$. Hence, there acts in $\hil$ a strongly-continuous unitary
representation $U$ of $\real^\dd$ which implements $\al$ i.e.
\beq
\pi(\al_x(A))=U(x)\pi(A)U(x)^{-1},\quad  A\in\mfa,\,\,\, x\in\real^\d.
\eeq   
We establish new relations between the spectral concepts on the side of $\al$ and $U$. In particular, we find
the following \emph{continuity transfer relations}
\beqa
 & &\pi(\mfa_{\cc})\Om\subset\hil_{\cc},\\ 
& &\pi(\mfa_{\ac})\Om\subset\hil_{\ac},
\eeqa
akin to the spectrum transfer property~(\ref{continuity-transfer}) from the standard Arveson theory.
We also verify the inclusions  
\beqa
& &\pm\Sp_{\sic}U\subset\Sp_{\sic}\al, \label{singular-continuous-two} \\
& &\pm\Sp_{\ac}U\subset\Sp_{\ac}\al^*, \label{absolutely-continuous-two}
\eeqa
which provide means to estimate the shapes of the spectra introduced above.

This  analysis applies, in particular, to the group of spacetime translation automorphisms $\al$
in any local, relativistic quantum field theory equipped with a normal vacuum state $\om_0$. We
obtain from inclusions~(\ref{singular-continuous-two}), (\ref{absolutely-continuous-two})  that in a theory of Wigner particles 
the singular-continuous spectrum of $\al$ contains the mass hyperboloids of these particles, whereas the multiparticle spectrum
contributes to $\Sp_{\ac}\al^*$. The subgroup of space translation automorphisms  $\be_{\vx}=\al_{(0,\vx)}$ allows for a more
detailed analysis.
Relying on results from \cite{Bu90}, we obtain that $\Sp_{\ac}\be=\Sp_{\ac}\be^{*}=\real^s$,
whereas $\Sp_{\sic}\be=\Sp_{\sic}\be^*$ are either  empty or consist of $\{0\}$. These spectra turn out to be empty in theories satisfying
certain timelike asymptotic abelianess condition introduced in \cite{BWa92} or complying with a regularity criterion~$\B$ which restricts
the continuous spectrum of $\al$ in a neighborhood of zero. 
We show, following~\cite{Dy08.3}, that this latter condition implies the existence of particles in theories containing a stress-energy tensor.

This paper is organized as follows: In Section~\ref{pp-section} 
we recall the basics of the Arveson spectral theory, introduce the continuous Arveson spectrum 
and decompose it into the absolutely continuous and singular-continuous parts.
Section~\ref{Dual-spectral} focuses on  relations between the spectra of $\al$ and $\al^*$.  
Unitarily implemented groups of automorphisms acting on $C^*$-algebras are studied
in Section~\ref{Algebraic-Setting}. In Section~\ref{QFT-section} we establish general properties of 
the continuous spectrum of spacetime translation automorphisms valid in any local relativistic quantum field theory  
admitting a normal vacuum state.
In Section~\ref{triviality-of-Apc}, which contains some results from author's PhD thesis \cite{Dy08.3}, we restrict attention to models complying with the regularity condition~$\B$. 
We strengthen our spectral results in  this setting and  show that
such theories describe particles, if they contain a stress-energy tensor. 
Section~\ref{section:conclusions} summarizes our results and outlines future directions. 
In the Appendix
we consider spectral theory of space translation automorphisms  in the absence of normal vacuum states. In particular, we
provide examples which violate  relations~(\ref{point+cont}), (\ref{point*+cont*}).

\section{Continuous spectrum of a group of isometries} \label{pp-section} 
\setcounter{equation}{0}

In this section we consider a strongly continuous group of isometries $\real^{\dd}\ni x\to\al_x$ 
acting on a Banach space $\mfa$. We choose a  subspace $\mfa_*\subset\mfa^*$ 
which satisfies the following: 
\begin{enumerate}
\item[] \bf Condition $\C$: \rm The subspace $\mfa_*$ is norm closed in $\mfa^*$ and invariant under the action of $\al^*$. 
Moreover, for any $A\in\mfa$,
\beq
\|A\|=\sup_{\fun\in\mfa_{*,1}}|\fun(A)|. \label{separation}
\eeq
\end{enumerate}
The foundations of  spectral theory in this setting were laid by Arveson~\cite{Ar74,Ar82}. 
We recall below the familiar concepts of spectral subspaces and  the pure-point spectrum.  
Next, we propose a new notion of continuous Arveson spectrum and decompose it into the absolutely continuous and
singular-continuous parts. For this latter purpose we  choose a norm dense subspace $\hmfa_{*}\subset\mfa_{*}$.

For any $\fun\in\mfa_*$ and $A\in\mfa$  we consider the Fourier transforms of  bounded, continuous functions 
$\real^d\ni x\to\fun(\al_x(A))$ 
\beq
\fun(\wt{A}(p)):=\fr{1}{(2\pi)^{\fr{d}{2}}}\int d^dx\, e^{-ip x}\fun(\al_{x}(A)), \label{Fourier-transform-convention}%
\eeq
which are tempered distributions. Here $px$ stands for some non-degenerate inner product in $\real^\dd$ and we adhere to the Fourier transform convention
which omits the minus sign in front of $px$ in the case  of test-functions. 
The Arveson spectrum of the group of isometries $\al$ is defined as follows
\beq
\Sp\,\al:=\ov{\underset{\su{A\in\mfa \\ \fun\in\mfa_*}}\bigcup\supp\,\fun(\wt{A}(\ccdot))}. \label{Arveson-spectrum}
\eeq
We also define the Arveson spectrum of an individual element $A\in\mfa$ as $\Sp^A\al:=\Sp\,\al|_{X}$, where 
$X=\Span\{\,\al_x(A) \,|\, x\in\real^d\}^{\no\cl}$ and $\no\cl$ denotes the norm closure.
(We mention as an aside that for a strongly continuous one-parameter group of isometries, $\Sp\,\al$ coincides with the operator-theoretic 
spectrum of the infinitesimal generator $D=\fr{1}{i}\fr{d}{dt}\al_t|_{t=0}$ \cite{Ev76, Lo77}).
Similarly as in the Hilbert space setting,  for any  closed set $\De\subset\real^\dd$ one defines the spectral subspace
\beq
\tmfa(\De):=\{\,A\in\mfa \,|\, \underset{\fun\in\mfa_*}\forall\supp\,\fun(\wt{A}(\ccdot))\subset \De\,\}.\label{spectral-subspace}
\eeq
In particular, for a single point $q\in\real^{\dd}$ there holds
\beq
\tmfa(\{q\})=\{\, A\in\mfa \,|\, \al_x(A)=e^{iqx}A \ \te{ for all } \ x\in\real^{\dd} \,\}.
\eeq
This leads us to  natural definitions of the pure-point subspace and the pure-point spectrum, 
analogous to those from the Hilbert space setting: 
\beqa
\mfa_{\spp}&:=&\Span\{\,\tmfa(\{q\})\,|\, q\in \Sp\,\al\,\}^{\no\cl}, \label{pure-point-subspace}\\
\Sp_{\spp}\al&:=&\{\, q\in \Sp\,\al \,|\, \tmfa(\{q\})\neq \{0\}\, \}.
\eeqa
We note that these objects are independent of the choice of $\mfa_*$, 
as long as it satisfies Condition~$\C$. 

The spectral subspaces (\ref{spectral-subspace}) and the pure-point spectrum have been
thoroughly studied in the  literature. (See \cite{Ar82,Pe} for a review of the former subject and \cite{Ba78,Hu99}
for interesting results on the latter). However, there does not seem to exist any generally accepted
definition of the continuous Arveson spectrum. Motivated by the Ergodic Theorem from the setting
of groups of unitaries acting on Hilbert spaces \cite{RS1}, we propose
 the following norm closed, invariant subspace $\mfa_{\co}$ and the corresponding spectrum
\beqa
\mfa_{\co}&:=&\{\, A\in\mfa\,|\, \underset{\su{q\in\real^\dd \\ \fun\in\mfa_*} }\forall \ 
\underset{K\nearrow \real^{\dd}}\lim\fr{1}{|K|}\int_{K} e^{-iqx}\fun(\al_x(A))\,d^{\dd}x=0\, \},\label{continuous-subspace}\\
\Sp_{\co}\al&:=&\Sp\,\al|_{\mfa_{\co}}. \label{co-Banach}
\eeqa
Here $K\nearrow \real^{\dd}$ denotes a family of cuboids, centered at zero, whose edge lengths tend to infinity\footnote{In some cases
we impose further restrictions on this family. See the discussion preceding Theorem~\ref{key-assumption-verification}.}.
In contrast to the pure-point part, the continuous subspace depends on the choice of 
$\mfa_*$. However,  for any such choice $\mfa_{\spp}\cap\mfa_{\cc}=\{0\}$, if $\Sp_{\spp}\al$ is a finite set. On the
other hand, the equality $\mfa=\mfa_{\spp}\oplus\mfa_{\scc}$, expected from the Hilbert space setting,
fails in some cases, as we show in the Appendix. Necessary and sufficient conditions for this equality, which we establish 
in Section~\ref{Dual-spectral}, will suggest judicious choices of $\mfa_*$.

Let us now introduce more refined spectral concepts which are sensitive to regularity
properties of the distributions $\fun(\wt{A}(\ccdot))$.
With the help of the norm dense subspace $\hmfa_{*}$ in $\mfa_{*}$ we define the absolutely continuous subspace
and the corresponding spectrum:
\beqa
\mfa_{\ac}&:=&\{\, A\in\mfa_{\cc} \,|\,\underset{\fun\in\hmfa_{*} }\forall \fun(\wt{A}(\,\cdot\,))\in L^1(\real^\dd,d^\dd p)  \,\}^{\no\cl},\label{ac-subspace}\\
\Sp_{\ac}\al&:=&\Sp\,\al|_{\mfa_{\ac}}.\label{ac-spectrum}
\eeqa
Let us recall that the singular-continuous part of a Hilbert space is the orthogonal
complement of the absolutely continuous subspace in the continuous one.
In the  case of a Banach space  there may not exist any direct sum complement of $\mfa_{\ac}$
which is invariant under the action of $\al$. (As a matter of fact, there may not exist any direct sum complement
at all \cite{Ru}). Therefore, we define the singular continuous \emph{space} as a quotient
\beq
\mfa_{\sic}:=\mfa_{\cc}/\mfa_{\ac}. \label{singular-subspace}
\eeq
We denote the equivalence class of an element $A\in\mfa_{\cc}$ by $[A]$ and introduce
the strongly continuous group of isometries $\real^\dd\ni x\to\alq_x$ of $\mfa_{\sic}$ given by
\beq
\alq_x[A]:=[\al_x(A)]. \label{singular-automorphisms}
\eeq
We define the singular continuous spectrum as the Arveson spectrum of $\alq$
\beq
\Sp_{\sic}\al:=\Sp\,\alq=\ov{\bigcup_{\su{A\in\mfa_{\cc}\\ \fun\in\mfa_{\sic}^*}}\supp\,\fun(\wt{[A]}(\ccdot))},
\label{singular-spectrum}
\eeq
where $\fun(\wt{[A]}(p))=(2\pi)^{-\fr{\dd}{2}}\int d^dx\, e^{-ip x}\fun(\alq_{x}([A]))$. Clearly, $\Sp_{\sic}\al=\emptyset$,  if and only if $\mfa_{\ac}=\mfa_{\cc}$. 

To conclude this section, let us consider briefly the case of a group of unitaries  $\al$ acting on a Hilbert space $\mfa$ 
with $\hmfa_*=\mfa_*=\mfa^*$. Noting that the distributions $\fun(\wt{A}(\ccdot))$ are then just the spectral measures, we obtain
that the above  concepts of the pure-point, continuous and absolutely continuous spectral
subspaces, and the corresponding spectra coincide with the standard ones. Moreover, $\Sp_{\sic}\al$ is equal to the conventional 
singular-continuous spectrum and the space~(\ref{singular-subspace}) is isomorphic, as a
Banach space, to the singular-continuous subspace.


\section{Spectral analysis of the transposed action  } \label{Dual-spectral} 

In the setting of groups of unitaries, acting on a Hilbert space, there always holds
\beq
\mfa=\mfa_{\spp}\oplus\mfa_{\cc}. \label{point+continuous}
\eeq
However, as shown in the Appendix, in the Banach space setting  
the above equality fails, if the  space $\mfa_{*}$ is excessively large. In order to characterize the choices of
$\mfa_*$ which entail relation~(\ref{point+continuous}), we develop the spectral theory for the transposed action $\al^*$.
Thus we define the spectra $\Sp\,\al^*$, $\Sp_{\spp}\al^*$, $\Sp_{\cc}\al^*$,  as well as the subspaces  $\tmfa_*(\De)$,
$\mfa_{*,\spp}$, $\mfa_{*,\cc}$ analogously as in the previous section, by exchanging  the roles of $\mfa_*$ and $\mfa$.
After selecting a norm dense subspace $\hmfa\subset\mfa$, we also define  $\mfa_{*,\ac}$,  
$\mfa_{*,\sic}$ and the corresponding spectra $\Sp_{\ac}\al^*$, $\Sp_{\sic}\al^*$\footnote{If $\un{\al^*}$ does not 
act norm continuously on $\mfa_{*,\sic}$, we refrain from defining $\Sp_{\sic}\al^*$. All the other definitions
remain meaningful, if the action of $\al^*$ is only  weakly$^*$ continuous.}.

The evaluation of functionals from $\mfa_*$ on elements of $\mfa$
provides a natural substitute for the scalar product from the Hilbert space setting. We exploit this observation in 
the following lemma. To stress the analogy with the Hilbert space case, we denote this evaluation by $\lan\,\cdot\,,\,\cdot\,\ran$. Moreover, for any subset $\mfa_{0}\subset\mfa$ we define the annihilator $\mfa_{0}^{\bot}=\{\, \fun\in\mfa_*\,|\,  \forall_{A\in\mfa_{0}}\,\, \fun(A)=0\,\}$, and analogously for subsets of~$\mfa_*$. 
\bel\label{orthogonal} There holds
\begin{enumerate}
\item[(a)] $\lan \tmfa_*(\De), \tmfa(\De^\prime)\ran=0$, if $\De,\De^\prime\subset\real^{\dd}$ are closed and $\De\cap\De^\prime=\emptyset$,
\item[(b)] $\lan\mfa_{*,\spp},\mfa_{\cc}\ran=0$ and $\lan\mfa_{*,\cc},\mfa_{\spp}\ran=0$.
\end{enumerate}
\eel
\proof  As for part (a), let $\chi_{\De}$, (resp. $\chi_{\De^\prime}$), be a bounded, smooth function on $\real^d$ which is equal to one on $\De$, 
(resp. on $\De^\prime$). Moreover, suppose  that $\supp\,\chi_{\De}\cap\supp\,\chi_{\De^\prime}=\emptyset$. 
Then there holds for any $\fun\in\tmfa_*(\De)$, $A\in\tmfa(\De^\prime)$ and  $f\in S(\real^d)$
\beqa
\int d^dx\,\fun(\al_x(A)) f(x)&=&\int d^dp\,\fun(\wt{A}(p))\tf(p)\non\\
&=&\int d^dp\,\fun(\wt{A}(p))\,\tf(p)\chi_{\De^\prime}(p)\,\chi_{\De}(p)=0.
\eeqa
Therefore 
$\real^d\ni x\to\fun(\al_x(A))$ vanishes as a distribution. Since it is a continuous function, it vanishes pointwise.

To prove (b), suppose that $\fun\in\mfa_*$ is an eigenvector i.e. $\al^*_x\fun=e^{iqx}\fun$ for some $q\in\real^\dd$. 
Then there holds for any $A\in\mfa_{\scc}$
\beq
\fun(A)=\fr{1}{|K|}\int_{K}d^dx\,e^{-iqx}e^{iqx}\fun(A)=\fr{1}{|K|}\int_{K}d^dx\,e^{-iqx} \fun(\al_x(A))\underset{K\nearrow\real^d}{\to}0.
\eeq    
Since eigenvectors form a total set in $\mfa_{*,\spp}$, the proof of the first statement in (b) is complete. The proof
of the second statement is analogous. \qed\\
With the help of the above observation  we easily obtain the following list of necessary conditions 
for the decomposition~(\ref{point+continuous}). 
\bet\label{main-ppc-theorem} Suppose that $\mfa=\mfa_{\spp}\oplus\mfa_{\cc}$. Then: 
\begin{enumerate}
\item[(a)] For any $q\in\real^\dd$ there holds $\dim\,\tmfa(\{q\})\geq \dim\,\tmfa_*(\{q\})$, 
\item[(b)] $\mfa^{\bot}_{\spp}=\mfa_{*,\cc}$,
\item[(c)] $\Spt_{\spp}\al\supset\Spt_{\spp}\al^*$ and  $\Spt_{\cc}\al\supset\Spt_{\cc}\al^*$.
\end{enumerate}
\eet
\nin\bf Remark. \rm  This result  also holds with the roles of $\mfa$ and $\mfa_*$ exchanged.\\

\proof In part (a) it suffices to consider the case $\dim\,\tmfa(\{q\})=m<\infty$. Suppose there
exist $m+1$ linearly independent functionals $\{\fun_k\}_1^{m+1}$ in $\tmfa_*(\{q\})$. 
Then, according to Lemma~\ref{orthogonal}~(b), their restrictions  $\hat{\fun}_k$ to the subspace
\beq
\mathring{\mfa}_{\spp}:=\Span\{\,\tmfa(\{q'\})\,|\, q'\in\Sp_{\spp}\al\,\} 
\eeq
which consists of finite, linear combinations of eigenvectors, still form a linearly independent family.
In view of Lemma~\ref{orthogonal}~(a), 
\beq
\ker\,\hat{\fun}_1\cap\ldots\cap\ker\,\hat{\fun}_{m+1}\supset \Span\{\,\tmfa(\{q^\prime\})\,|\, q^{\prime}\in\Sp_{\spp}\al,\, q^\prime\neq q\,\},
\eeq 
which is a contradiction, since the subspace on the r.h.s. of this relation has codimension $m$ in $\mathring{\mfa}_{\spp}$. 

As for part (b), we note that $\mfa^{\bot}_{\spp}\supset\mfa_{*,\cc}$, by Lemma~\ref{orthogonal}~(b). In
order to prove the opposite inclusion, we pick $\fun\in\mfa^{\bot}_{\spp}$ and arbitrary $A\in\mfa$. According to
our assumption, $A=A_{\spp}+A_{\cc}$, where $A_{\spp}\in\mfa_{\spp}$ and $A_{\cc}\in\mfa_{\cc}$. There holds for
any $q\in \real^\dd$ 
\beq 
\fr{1}{|K|}\int_{K}d^\dd x\,e^{-iqx} \fun(\al_x(A))=\fr{1}{|K|}\int_{K}d^dx\,e^{-iqx} \fun(\al_x(A_{\cc}))\underset{K\nearrow\real^d}{\to}0,
\eeq
and therefore $\fun\in\mfa_{*,\cc}$. 

The statement concerning the point spectrum in (c) follows immediately from part (a). As for the continuous spectrum, we note that
\beq
\Sp_{\cc}\al=\ov{\underset{\su{A\in\mfa_{\cc} \\ \fun\in\mfa_*}}\bigcup\supp\,\fun(\wt{A}(\ccdot))}
\supset \ov{\underset{\su{A\in\mfa_{\cc} \\ \fun\in\mfa_{*,\cc}}}\bigcup\supp\,\fun(\wt{A}(\ccdot))}
=\Sp_{\cc}\al^*,
\eeq
where the last equality relies on the fact that $\mfa^{\bot}_{\spp}\supset\mfa_{*,\cc}$ and on the assumption. \qed\\
In the Appendix we   find counterexamples to relation~(\ref{point+continuous}). To this end, we exploit part (a) of the above theorem as follows: 
First, we show  that for the group of space translation automorphisms acting on the algebra of observables $\mfa$ in  quantum field theory 
there always holds $\dim\,\tmfa(\{0\})=1$. Next, we note that $\tmfa_*(\{0\})$ is spanned by the functionals which are invariant under the 
transposed action. Hence $\mfa\neq\mfa_{\spp}\oplus\mfa_{\cc}$, if  $\mfa_*$ contains more than one vacuum state\footnote{That is a translationally 
invariant state on the algebra of observables s.t. the relativistic spectrum condition holds in its GNS representation (cf. Section~\ref{QFT-section}). }.

Coming back to the general setting of groups of isometries, we provide sufficient conditions for relation~(\ref{point+continuous}) to hold.
The following theorem accounts for the important role of the point spectrum of $\al^*$ in the study of the continuous spectrum of $\al$.
Its assumptions  have a natural formulation in the framework of automorphism groups of $C^*$-algebras which we explore in the next section. 
\bet\label{vacuum-kernel} Suppose there exists a functional $\om_0\in\mfa_*$ s.t. $\ker\om_0\subset\mfa_{\cc}$ and a non-zero element 
$I\in\tmfa(\{q\})$ for some $q\in\real^\dd$.
Then: 
\begin{enumerate} 
\item[(a)] $\mfa=\mfa_{\spp}\oplus\mfa_{\cc}$, where $\mfa_{\spp}=\Spant\,\{I\}$,   $\mfa_{\cc}=\ker\,\om_0$,
\item[(b)] $\om_0\in\tmfa_*(\{q\})$,
\item[(c)] $\mfa_*=\mfa_{*,\spp}\oplus\mfa_{*,\cc}$, where $\mfa_{*,\spp}=\Spant\,\{\om_0\}$, $\mfa_{*,\cc}=\ker\, I$.
\end{enumerate}
Moreover, $\Spt_{\cc}\al=\Spt_{\cc}\al^*$.
\eet
\proof By assumption, $\mfa_{\cc}$ is a subspace of codimension at most one in $\mfa$. Since $\mfa_{\spp}\supset\Span\,\{I\}$, 
part (a) follows. 
Hence any $A\in\mfa$ can be expressed as
$A=cI+A_{\cc}$, where $A_{\cc}\in\mfa_{\cc}$. Since $\mfa_{\cc}$ is invariant under the action of automorphisms, there holds
$\al^*_x\om_0(A)=c\om_0(\al_x(I))=e^{iqx}\om_0(A)$, which entails (b). From Theorem~\ref{main-ppc-theorem}~(b) 
we obtain that $\mfa_{*,\cc}=\ker\, I$.
Since it is a subspace of codimension one, part (c) follows from part (b). The last statement is a consequence of  
Theorem~\ref{main-ppc-theorem}~(c) and parts (a), (c) of the present theorem. \qed


\section{Spectral analysis of a group of automorphisms} \label{Algebraic-Setting}

In this section we consider a group of automorphisms $\al$ acting on a  
$C^*$-algebra $\mfa$ containing a unity $I$.  We assume that there exists a pure state $\om_0$ on
$\mfa$ which satisfies 
\beq
\ker\,\om_0\subset \mfa_{\cc}, \label{key-assumption}
\eeq
where $\mfa_*$, entering the definition of $\mfa_{\cc}$, is chosen as the predual of the GNS representation  $(\pi,\hil,\Om)$ induced
by the state $\om_0$. (As we will see in Section~\ref{QFT-section},
in the case of spacetime translation automorphisms in QFT any pure vacuum state satisfies this inclusion as a
consequence of locality). On the subspace $\hmfa_*\subset\mfa_*$ we impose the following condition:
\begin{enumerate}
\item[] \bf Condition $\DC$: \rm The subspace $\hmfa_{*}$ is    
self-adjoint\footnote{That is if $\fun\in\hmfa_*$ then $\ov{\fun}\in\hmfa_{*}$, where $\ov{\fun}(A)=\ov{\fun(A^*)}$, $A\in\mfa$.}, 
norm dense in $\mfa_{*}$ and contains all the functionals of the form $(\Psi|\pi(\,\cdot\,)\Om)$, where $\Psi$ belongs to some dense subspace in $\hil$. 
\end{enumerate}
Having specified the framework, we proceed to the spectral analysis of $\al$. 
Exploiting Theorem~\ref{vacuum-kernel} and the fact that $\mfa$ is unital, we obtain:
\bet\label{pp-c-decomposition} Let $\mfa$, $\om_0$ and $\mfa_*$ be specified as above. Then: 
\begin{enumerate} 
\item[(a)] $\mfa=\mfa_{\spp}\oplus\mfa_{\cc}$, where $\mfa_{\spp}=\Spant\,\{I\}$,   $\mfa_{\cc}=\ker\,\om_0$,
\item[(b)] $\om_0\in\tmfa_*(\{0\})$,
\item[(c)] $\mfa_*=\mfa_{*,\spp}\oplus\mfa_{*,\cc}$, where $\mfa_{*,\spp}=\Spant\,\{\om_0\}$, $\mfa_{*,\cc}=\ker\, I$.
\end{enumerate}
Moreover, $\Spt_{\cc}\al=\Spt_{\cc}\al^*$.
\eet
\nin In view of part (b) of the above theorem, the state $\om_0$ is invariant under the action
of $\al^*$. Hence there exists a strongly continuous group of unitaries $\real^d\ni x\to U(x)$, 
acting on $\hil$, s.t. $\pi(\al_x(A))=U(x)\pi(A)U(x)^{-1}$ and the vector
$\Om\in\hil$ is invariant under its action. We denote by $\thil(\De)$, $\hil_{\spp}$, $\hil_{\cc}$, $\hil_{\ac}$, $\hil_{\sic}$
the spectral subspaces of $\hil$ w.r.t. the action of $U$. 
One of the central  problems in the present setting is to find  relations
between the Arveson spectrum of $\al$ and the spectrum of the implementing group of unitaries. An important 
and well known property is the \emph{spectrum transfer}: If $A\in\tmfa(\De_1)$ then
\beq
\pi(A)\thil(\De_2)\subset \thil(\ov{\De_1+\De_2}),\label{continuity-transfer}
\eeq  
where $\De_1,\De_2\subset\real^{\dd}$ are closed sets \cite{Ar82}. 
It turns out that similar properties, which can be called \emph{continuity transfer} relations, hold at the level of more
detailed spectral theory: 
\bep\label{pc-hil} Let $\mfa$, $\om_0$ and $\mfa_*$ be specified as above and suppose that Condition~$\DC$ holds. Then: 
\begin{enumerate}
\item[(a)] $\hil_{\spp}=\Spant\{\Om\}$,
\item[(b)] $\hil_{\cc}=\{\, \pi(\mfa_{\cc})\Om\,\}^{\no\cl}$,  
\item[(c)] $\hil_{\ac}\supset \pi(\mfa_{\ac})\Om$.
\end{enumerate}
\eep
\proof First, we show that $\pi(\mfa_{\cc})\Om\,\subset\hil_{\cc}$: Let $A\in\mfa_{\cc}$ and $\Psi\in\hil$ be arbitrary. 
Then  for any $q\in\real^{\dd}$
\beq
\fr{1}{|K|}\int_{K}d^dx\,e^{-iqx} (\Psi|U(x)\pi(A)\Om)=\fr{1}{|K|}\int_{K}d^dx\,e^{-iqx} \fun(\al_x(A))
\underset{K\nearrow\real^{\dd}}{\to} 0,
\eeq
where in the last step we made use of the fact that $\fun(\,\cdot\,)=(\Psi|\pi(\,\cdot\,)\Om)\in\mfa_{*}$. Thus, by the Ergodic Theorem, $\pi(A)\Om\in \hil_{\cc}$.

Next, we verify that $\{\, \pi(\mfa_{\cc})\Om\,\}^{\no\cl}=\{\Om\}^{\bot}$. Suppose that $\Psi\in\hil$ and $(\Psi|\Om)=0$. 
By cyclicity of $\Om$, there
exists a sequence $\{A_n\}_{n\in\nat}$ of elements of $\mfa$ s.t. $\|\Psi-\pi(A_n)\Om\|\to 0$. Then
$B_n:=A_n-\om_0(A_n)I$ is a sequence of elements of $\mfa_{\cc}$ which 
satisfies $\|\Psi-\pi(B_n)\Om\|\to 0$. This completes the proof of (a) and (b). 

To prove (c), let $A\in\mfa_{\ac}$ and let $\{A_n\}_{n\in\nat}$  be a sequence of elements of $\mfa_{\ac}$
which converges to $A$ in norm and s.t.
\beq
\int d^\dd p\,|\fun(\wt{A}_n(p))|<\infty
\eeq 
for all $n\in\nat$ and $\fun\in\hmfa_{*}$. Then, for any $\Psi$ from the dense set appearing in Condition~$\DC$,  
\beq
\int d^\dd p\,|\wt{(\Psi|U(\,\cdot\,)\pi(A_n)\Om)}(p)|=\int d^\dd p\,|(\Psi|\pi(\wt{A}_n(p))\Om)|<\infty.
\eeq 
Hence the  vectors $\pi(A_n)\Om$ belong to $\hil_{\ac}$ and so does their norm limit $\pi(A)\Om$. \qed\\
After this preparation we establish relations between the spectra of $U$ and $\al$. We use these facts in
the next section,  in a quantum field theoretic context. 
\bet\label{continuous-proposition}  Under the assumptions of Proposition~\ref{pc-hil} there hold the relations:
\begin{enumerate}
\item[(a)] $\Spt_{\spp}U=\Spt_{\spp}\al=\Spt_{\spp}\al^{*}=\{0\}$,  
\item[(b)] $\Spt_{\cc}U-\Spt_{\cc}U\subset \Spt_{\cc}\al$, 
\item[(c)] $\pm\Spt_{\ac}U\subset \Spt_{\ac}\al^*$,
\item[(d)]  $\pm\Spt_{\sic}U\subset \Spt_{\sic}\al$.
\end{enumerate}
\eet
\proof Part (a) follows immediately from Proposition~\ref{pc-hil}~(a) and Theorem~\ref{pp-c-decomposition}.
To prove (b), let $q=p_1-p_2$, where $p_1,p_2\in\Sp_{\cc}\,U$. We choose an open neighborhood $V_q$  of 
$q$ and bounded neighborhoods $\De_1$, $\De_2$  of $p_1$ and $p_2$, respectively,  s.t. $\ov{\De_1-\De_2}\subset V_q$.
We pick $\Psi_1\in \Ran\,P(\De_1)$ and $\Psi_2\in\Ran\,P(\De_2)$. Since the continuous spectrum of
$U$ is a subset of the essential spectrum, we can find such $\Psi_1$, $\Psi_2$ that $(\Psi_1|\Psi_2)=0$. 
By purity of the state $\om_0$, $\pi(\mfa)$ acts irreducibly on $\hil$ and we can find such $A\in\mfa$ that  
\beq
(\Psi_1|\pi(A)\Psi_2)\neq 0.
\eeq
Replacing $A$ with $A-\om_0(A)I$, if necessary, we can assume that $A\in\mfa_{\cc}$. Now we choose $f\in S(\real^{\dd})$
s.t. $\tf(p)=(2\pi)^{-\dd/2}$ for  $p\in\ov{\De_1-\De_2}$ and $\tf(p)=0$ for $p$ outside of $V_q$. 
Making use of the fact that the support of the distribution  $(\Psi_1|\pi(\wt{A}(\ccdot))\Psi_2)$ is 
contained in $\ov{\De_1-\De_2}$, which follows from relation~(\ref{continuity-transfer}), we obtain
\beqa
\int\,d^\dd x\, f(x)(\Psi_1|\pi(\al_x(A))\Psi_2)=\int\,d^\dd p\, \tf(p)(\Psi_1|\pi(\wt{A}(p))\Psi_2)& &\non\\ 
=(\Psi_1|\pi(A)\Psi_2)\neq 0,& &        
\eeqa
which implies that $q\in\Sp_{\cc}\al$.

As for (c), let $q\in \Sp_{\ac} U$ and let $V_q$ be any open neighborhood of $q$. Then there exists  $f\in S(\real^{\dd})$ s.t. $\supp\,\tf\subset V_q$ and $\Psi_f:=\int d^\dd x\, f(x)U(x)\Psi\neq 0$ for some $\Psi\in\hil_{\ac}$. We note that the functional 
$\fun_{\Psi}(\,\cdot\,):=(\Psi|\pi(\,\cdot\,)\vac)$  belongs to $\mfa_{*,\ac}$, since for any
$A\in\mfa$ 
\beq
\int d^{\dd}p\,|\fun_{\Psi}(\wt{A}(p))|=\int d^{\dd}p\, |\wt{(\Psi|U(\,\cdot\,)\pi(A)\vac)}(p)|<\infty,
\eeq 
where in the last step we made use of the fact that $\Psi\in\hil_{\ac}$. Next, since $\pi(\mfa)\Om$ is dense
in $\hil$, we find such $B\in\mfa$ that $(\Psi_f|\pi(B)\Om)\neq 0$. Hence
\beq
\int d^\dd x\, \bar f_-(x)\,\al^*_x\fun_{\Psi}(B)\neq 0,
\eeq
where $f_-(x)=f(-x)$. Since $\supp\,\tilde{\bar f}_-\subset V_q$, we obtain that $q\in\Sp_{\ac}\al^*$. 
To show that $-q\in\Sp_{\ac}\al^*$, we repeat the argument using the functional
$\ov{\fun}_{\Psi}(\,\cdot\,)=(\vac|\pi(\,\cdot\,)\Psi)$ instead of $\fun_{\Psi}$. 

To prove (d), let $q\in \Sp_{\sic} U$ and let $V_q$ be an open neighborhood of $q$. Then there exists  $f\in S(\real^{\dd})$ s.t. $\supp\,\tf\subset V_q$ and $\Phi_f:=\int d^\dd x\, f(x)U(x)\Phi\neq 0$ for some $\Phi\in\hil_{\sic}$. By Proposition~\ref{pc-hil}~(b), $\pi(\mfa_{\cc})\Om$ is dense in $\hil_{\cc}$, so we can find $A\in\mfa_{\cc}$ s.t. 
\beq
(\Phi_f|\pi(A)\vac)\neq 0. \label{non-triviality-A} 
\eeq
We note that the functional $\fun_{\Phi}(\,\cdot\,)=(\Phi|\pi(\,\cdot\,)\vac)$    
is an element of $\mfa_{\cc}^*$ which, by Proposition~\ref{pc-hil}~(c), contains $\mfa_{\ac}$ in its kernel. 
Therefore, $\fun_{\Phi}$ induces a well defined, bounded functional $\funq_{\Phi}$ on $\mfa_{\sic}=\mfa_{\cc}/\mfa_{\ac}$ s.t.
$\funq_{\Phi}([B])=\fun_{\Phi}(B)$ for $B\in\mfa_{\cc}$. By relation~(\ref{non-triviality-A}), 
\beq
\int d^\dd x\, \bar f_-(x)\funq_{\Phi}(\alq_x[A])\neq 0,
\eeq
which proves that  $q\in\Sp_{\sic}\al$. To show that $-q\in\Sp_{\sic}\al$, we repeat the argument using the functional $\ov{\fun}_{\Phi}(\,\cdot\,)=(\vac|\pi(\,\cdot\,)\Phi)$ instead of $\fun_{\Phi}$.
\qed\\


\section{Spectral theory of automorphism groups in QFT}\label{QFT-section}  

\setcounter{equation}{0}

In this section we analyze the spectrum of the group of spacetime translation automorphisms acting on the algebra of
observables  in  quantum field theory. To keep our investigation general, we  rely on the  
Haag-Kastler framework of algebraic quantum field theory \cite{Ha}:
The theory is based on a net $\mco\to\mfa(\mco)\subset B(\hil)$ of unital $C^*$-algebras, attached to open,
bounded regions of spacetime $\mco\subset\real^{s+1}$, which satisfies isotony and locality. The $*$-algebra of
local operators is given by 
\beq
\lmfa:=\bigcup_{\mco\subset\real^{s+1}}\mfa(\mco) \label{local-operators} 
\eeq
and its norm closure $\mfa$ is irreducibly represented on the infinitely dimensional Hilbert space $\hil$.  Moreover, $\hil$ carries a strongly continuous unitary representation of translations $\real^{s+1}\ni x\to U(x)$ which satisfies the relativistic spectral condition i.e. the joint spectrum of the generators $H, P_1,\ldots, P_s$   is contained in the closed forward lightcone~$\cone$. 
It is also assumed that the translation automorphisms $\al_x(\,\cdot\,)=U(x)\,\cdot\, U(x)^*$ act geometrically on the net i.e.
\beq
\al_{x}(\mfa(\mco))=\mfa(\mco+x), 
\eeq
and are strongly continuous, that is any function $\real^{s+1}\ni x\to\al_{x}(A)$, $A\in\mfa$, 
is continuous in the norm topology of $\mfa$. Suppose that the Hilbert space contains a vacuum vector $\Om$, which is invariant
under the action of $U$. Then we set $\om_0(\ccdot)=(\Om|\ccdot\Om)$ and say that the theory admits a normal vacuum state $\om_0$. 
This state belongs to the general class of functionals of bounded energy, defined as follows:
Let $P_E$ be the spectral projection of $H$ (the Hamiltonian) on the subspace spanned by vectors of energy lower than or 
equal to $E$. We identify  $B(\hil)_*$ with the space of trace-class operators on $\hil$ and denote by $S_E$ the set of states
from  $P_E B(\hil)_* P_E$. The following norm dense subspace of $B(\hil)_*$ 
\beq
\tracabd:=\Span\{\bigcup_{E\geq 0}S_E\} \label{bounded-energy}
\eeq
is called the space of functionals of bounded energy.

It is easy to incorporate the triple $(\al,\mfa,\om_0)$, introduced above,
into the algebraic setting of the previous section: We note that the GNS representation of $\mfa$, 
induced by $\om_0$, coincides with the defining representation, up to unitary equivalence. We set 
\beqa
\mfa_*=B(\hil)_*,\quad \hmfa_*=\tracabd,\quad \hmfa=\lmfa    \label{spectral-triple}
\eeqa
and check that Condition~$\DC$ is satisfied. 
It remains to verify that there holds the key assumption~(\ref{key-assumption}) i.e. $\ker\,\om_0\subset \mfa_{\cc}$. To this end, we choose the family of sets $K\nearrow\real^{s+1}$, appearing in definition~(\ref{continuous-subspace}), so as to exploit the local structure of the theory. In the case of spacetime translations we set $K_L=[-L^\eps, L^{\eps}]\times [-L,L]^{\times s}$ for some $0<\eps<1$ and any $L\in\real_+$, while for the subgroup of space translations we choose  $K_L'= [-L,L]^{\times s}$. Clearly, $K_L\nearrow\real^{s+1}$ and $K_L'\nearrow\real^s$ as $L\to\infty$.
Since we work with the
Minkowski space $\real^{s+1}$,  we set $px=p_0x_0-\vep\vx$,   $\vep\vx=\sum_{j=1}^s p_jx_j$, in definitions~(\ref{Fourier-transform-convention}) 
and (\ref{continuous-subspace}).  Accordingly, $px=p_0x_0$ for the time axis and $px=-\vep\vx$  for the spacelike hyperplane~$\real^s$.
After this preparation, we obtain:
\bet\label{key-assumption-verification} Let $\real^{s+1}\ni x\to\al_x$ be the group of spacetime translation automorphisms acting on the algebra of 
observables $\mfa$ in a quantum field theory admitting a normal vacuum state $\om_0$ and let $\mfa_*=B(\hil)_*$. 
Then $\ker\om_0\subset\mfa_{\cc}$ and, consequently,
\begin{enumerate} 
\item[(a)] $\mfa=\mfa_{\spp}\oplus\mfa_{\cc}$, where $\mfa_{\spp}=\Spant\,\{I\}$,   $\mfa_{\cc}=\ker\,\om_0$,
\item[(b)] $\mfa_*=\mfa_{*,\spp}\oplus\mfa_{*,\cc}$, where $\mfa_{*,\spp}=\Spant\,\{\om_0\}$, $\mfa_{*,\cc}=\ker\, I$,
\item[(c)] $\Spt_{\cc}\al=\Spt_{\cc}\al^*$.
\end{enumerate}
The above statements are also true for the subgroup $\real^s\ni\vx\to\be_{\vx}:=\al_{(0,\vx)}$ of
space translation automorphisms. 
\eet
\proof  To verify that $\ker\,\om_0\subset\mfa_{\cc}$, we have to show that for 
any $A\in\ker\om_0$ and $q\in\real^{s+1}$ 
\beq
\te{w$^*$-}\lim_{L\to\infty}\fr{1}{|K_L|}\int_{K_L} d^{s+1}x\,e^{-iqx}\al_{x}(A)=0.\label{to-prove}
\eeq
This can be proven by a rather standard argument:
We note that for any normal functional $\fun\in\trace$ the function 
$\real^{s+1}\ni x\to\al_{x}^*\fun$ is continuous w.r.t. the norm topology in $\trace$ 
so the following Bochner integrals 
\beq
\fun_{L}=\fr{1}{|K_L|}\int_{K_L} d^{s+1}x\,e^{-iqx}\al^*_{x}\fun
\eeq
define functionals from $\trace$. Now we fix a state $\om\in\trace$  and obtain, from the Banach-Alaoglu theorem, a net 
$\{\,K_{L_\be}\subset\real^{s+1}\,|\, \be\in\mathbb{I}\,\}$ and a functional $\ovomi\in\mfa^*$ s.t.
\beq
\te{w$^*$-}\lim_{\be}\om_{L_\be}=\ovomi. 
\eeq
Next, for any $A\in\mfa$, we consider the following net of elements from $B(\hil)$
\beq
P_{L}(A):=\fr{1}{|K_L|}\int_{K_L}d^{s+1}x\, e^{-iqx}\,\al_{x}(A).
\eeq
By locality and the slow growth  of the timelike dimension of $K_L$, the net $\{P_{L}(A)\}_{L>0}$ satisfies 
\beq
\lim_{L\to\infty}\|[P_{L}(A),B]\|=0
\eeq
for any $B\in\mfa$. Therefore, all its limit points w.r.t. the weak$^*$ topology
on $B(\hil)$ are multiples of the identity  by the assumed irreducibility of $\mfa$. 
It follows that for any $\fun\in\trace$, $A\in\mfa$
\beq
\lim_{L\to\infty}\big(\fun_{L}(A)-\om_{L}(A)\fun(I)\big)=0. \label{V-invariance}
\eeq
Consequently
\beq
\te{w$^*$-}\lim_{\be}\fr{1}{|K_{L_\be}|}\int_{K_{L_\be}} d^{s+1}x\,e^{-iqx}\al_{x}(A)=\ovomi(A)I.
\eeq
By evaluating this relation on the translationally invariant state $\om_0$, we obtain that $\ovomi=\om_0$
for $q=0$ and $\ovomi=0$ otherwise, which entails equality~(\ref{to-prove}). Now parts (a), (b) and (c) follow 
from Theorem~\ref{pp-c-decomposition}.  By an obvious
modification of the above argument  we obtain the statement concerning the subgroup of space translations. \qed\\
Proceeding to more detailed analysis of the spectrum of $\al$, we obtain, with the help of Theorem~\ref{continuous-proposition},
the following facts: 
\bet\label{continuous-proposition-one}  Let $\real^{s+1}\ni x\to\al_x$ be the group of spacetime translation automorphisms acting on the algebra of 
observables $\mfa$ in a quantum field theory admitting a normal vacuum state $\om_0$. 
Let $m>0$ and suppose that 
$\Spt\,U=\{0\}\cup \HH_m\cup \GG_m$, where $\HH_m=\{\, p\in\real^{s+1}\,|\, p^2=m^2, p^0>0\,\}$ is the mass hyperboloid and
$\GG_m=\{\,p\in\real^{s+1}\,|\, p^2\geq (2m)^2, p^0>0\,\}$ is the multiparticle spectrum. 
Then there holds:
\begin{enumerate}
\item[(a)] $\Spt_{\spp}\al=\Spt_{\spp}\al^{*} =\{0\}$,  
\item[(b)] $\Spt_{\cc}\al=\Spt_{\cc}\al^{*}=\real^{s+1}$, 
\item[(c)] $\Spt_{\ac}\al^*\supset \pm \GG_m $,
\item[(d)] $\Spt_{\sic}\al\supset \pm \HH_m$.
\end{enumerate}
Here the subspaces $\mfa_*$, $\hmfa_*$ and $\hmfa$ are chosen as in (\ref{spectral-triple}).
\eet
\proof Parts (a) and (b) follow directly from the corresponding statements in Theorem~\ref{continuous-proposition} 
and from Theorem~\ref{key-assumption-verification}~(c). 
Part (c) is a consequence of Theorem~\ref{continuous-proposition}~(c) under the premise that $\GG_m \subset \Spt_{\ac}U$. To prove this fact, we
proceed as follows:
Let $\hil_m$ be the spectral subspace
corresponding to $\HH_m$ and let $\hil_0=\Ga(\hil_m)$ be the symmetric Fock space over $\hil_m$. Moreover, 
let $U_m$ be the restriction of $U$ to $\hil_m$ and let $U_0=\Ga(U_m)$ be its second quantization acting on $\hil_0$. 
Then, by the Haag-Ruelle scattering theory \cite{BF82, Ar,Dy05}, there exists the isometric wave-operator $\Om_+:\hil_0\to\hil$ which 
satisfies
\beq
\Om_+U_0(x)=U(x)\Om_+,\quad x\in\real^{s+1}. \label{intertwining}
\eeq         
It is a simple exercise to show that $\GG_m=\Sp_{\ac}U_0$. Hence, if $q\in \GG_m$, then for any open neighborhood $V_q$ of $q$
there exists $f\in S(\real^{s+1})$ s.t. $\supp\,\tf\subset V_q$ and
\beq
\int d^{s+1}x\, U_0(x)f(x)\Psi_0\neq 0
\eeq    
for some  $\Psi_0\in\hil_{0,\ac}$. 
Thus, making use of property~(\ref{intertwining}) and the fact that $\Om_+$ is an isometry, 
we obtain that $\Om_+\Psi_0\in\hil_{\ac}$ and
\beq
\int d^{s+1}x\, U(x)f(x)\Om_+\Psi_0\neq 0,
\eeq
which proves that $q\in \Sp_{\ac}U$.  

To justify (d), we note that $\HH_m$ is a set of Lebesgue measure zero in $\real^{s+1}$. Hence, if $q\in \HH_m$, then
it either belongs to the singular-continuous spectrum or to the pure-point spectrum of $U$. The latter possibility
is excluded by  Proposition~\ref{pc-hil}~(a) (or  Lemma~3.2.5 of \cite{Ha}). Now the statement follows from Theorem~\ref{continuous-proposition}~(d). \qed\\
This theorem exhibits an interplay between the spectral properties of $\al$  
and the particle aspects of quantum field theory: The mass hyperboloids of Wigner particles contribute to the singular-continuous
spectrum of $\al$. 
Thereby, this result  provides a large class of physically relevant examples 
of automorphism groups with non-empty singular-continuous spectrum. However, it leaves open the question of non-triviality
of $\Sp_{\ac}\al$.   
Since we do not have sufficient control over the full group of spacetime translations, we will study this problem in the case of
the subgroup $\real^s\ni\vx\to\be_{\vx}:=\al_{(0,\vx)}$ of space translation automorphisms.

Theorem~\ref{key-assumption-verification} fixes the decomposition of $\mfa$ and $\mfa_*$ into the pure-point and  continuous parts  
w.r.t. the action of $\be$.  To facilitate further analysis, we introduce the following  subspaces
\beqa 
& &\hmfa_{\cc}:=\{\, A\in\hmfa\,|\,\om_0(A)=0\,\},\label{hmfacc} \\
& &\hmfa_{*,\cc}:=\{\,\fun\in\hmfa_*\,|\, \fun(I)=0\,\} \label{hmfascc}
\eeqa 
which are norm dense in $\mfa_{\cc}$ and $\mfa_{*,\cc}$, respectively.
Our study of the absolutely continuous and singular-continuous spectrum of $\be$ is based on two ingredients: 
First of them  is the following fact, mentioned in \cite{Bu90}.
\bel\label{auxiliary} 
For any non-zero  $A\in\hmfa_{\cc}$ 
\beq
\ov{\bigcup_{\fun\in\hmfa_{*,\cc}} \te{\emph{supp}}\,\fun(\wt{A}(\ccdot))}=\real^s,\label{operator-support}
\eeq
where the Fourier transform is taken w.r.t. the group of space translations $\be$.
\eel
\proof Let $X$ be the set on the l.h.s. of relation~(\ref{operator-support}) and $X_0$ its
counterpart  with $\hmfa_{*,\cc}$ replaced with its norm closure $\mfa_{*,\cc}$. First, we prove 
that $X_0=\real^s$. In fact,
suppose that the complement of $X_0$ is a non-empty (open) set. Then, for any  operator $B\in\hmfa$, and $\fun\in\trace$,
the functional $\fun(\,\cdot \, B)-\fun(B\,\cdot \,)$ is in $\mfa_{*,\cc}$. Hence the following analytic function of $\vep\in\real^s$
\beq
\fun([\wt{A}(\vep),B])=\fr{1}{(2\pi)^{s/2}}\int\,d^sx\,e^{i\vep\vx}\fun([\be_{\vx}(A), B])
\eeq
is identically equal to zero. Thus $A$ belongs to the commutant of $\mfa$ which consists of multiples of the identity.
Since $\om_0(A)=0$, we obtain that $A=0$, which is a contradiction. 

Now suppose that $X$ has a non-empty (open) complement $O$ in $\real^s$. Then, for any $f\in S(\real^s)$ 
s.t. $\supp\,\tf\subset O$ and for any $\fun\in\hmfa_{*,\cc}$, $\fun(A(f))=0$ holds, where
\beq
A(f):=\int d^sx\, f(\vx)\be_{\vx}(A)\in\mfa. 
\eeq
Since $\hmfa_{*,\cc}$ is norm dense in $\mfa_{*,\cc}$, the same holds for $\fun\in\mfa_{*,\cc}$, contradicting
the fact that $X_0=\real^s$. \qed\\ 
The second ingredient is the following estimate, due to Buchholz \cite{Bu90}, 
\beq
\sup_{\su{\om\in S_E}}\int d^sp\,|\vep|^{s+1+\eps} |\om(\wt{A}(\vep))|^2<\infty, \label{2regularity}
\eeq
valid for any $E\geq 0$, any local observable $A\in\lmfa$ and $\eps>0$. With these two facts at hand, we are ready to analyze 
the  spectrum of $\be$.
\bet\label{refined-theorem-old} Let $\real^s\ni\vx\to\be_{\vx}$ be the group of space translation automorphisms acting on the algebra of 
observables $\mfa$ in a quantum field theory admitting a normal vacuum state $\om_0$. Then there holds:  
\begin{enumerate}
\item[(a)] $\Spt_{\spp}\be=\Spt_{\spp}\be^{*}=\{0\}$,  
\item[(b)] $\Spt_{\ac}\be=\Spt_{\ac}\be^*=\real^s$,
\item[(c)] $\Spt_{\sic}\be\subset\{0\}$, $\Spt_{\sic}\be^*\subset\{0\}$.   
\end{enumerate}
The subspaces $\mfa_*$, $\hmfa_*$ and $\hmfa$ are given by (\ref{spectral-triple}).
\eet
\proof Part (a) follows directly from Theorem \ref{key-assumption-verification}. 
To prove (b) and (c), we proceed as follows:
For any function $f\in C_0^{\infty}(\real^s)$ and $n\in\nat$ we introduce $f_n\in C_0^{\infty}(\real^s)$ given by
$\tf_n(\vep)=\tf(\vep)|\vep|^{2n}$. Next, for any $A\in\hmfa_{\cc}$ and $\fun\in\hmfa_{*,\cc}$ we set
\beqa
A(f_n)&:=&\int d^sx\,f_n(\vx)\, \be_{\vx}(A),\\
\fun_{f_n}&:=&\int d^sx\, f_n(\vx)\, \be^*_{\vx}\fun.
\eeqa
We note that $A(f_n)\in\hmfa_{\cc}$, since each local algebra $\mfa(\mco)$ is a norm closed subspace of $\mfa$ and the action of $\be$ 
is strongly continuous.  Similarly, $\fun_{f_n}\in\hmfa_{*,\cc}$, since $\be^*$ acts strongly continuously on $\trace$ and each $\fun\in\hmfa_{*,\cc}$ belongs to the closed subspace $\Span\,S_E\subset\trace$ for $E$  sufficiently large. 
Setting $4n>s+1$ and noting that $\wt{A(f_n)}(\vep)=(2\pi)^{s/2}\tf_n(\vep)\wt{A}(\vep)$, we obtain from estimate~(\ref{2regularity})
\beq
\underset{\su{\fun\in\hmfa_{*,\cc} \\  A\in\hmfa_{\cc}}}{\forall}\int d^sp\, |\fun(\wt{A(f_n)}(\vep))|^2<\infty.
\eeq
Hence, recalling  definition~(\ref{ac-subspace}) and noting that the distributions $\fun(\wt{A(f_n)}(\,\cdot\,))$ are compactly supported, 
we conclude that $A(f_n)\in\mfa_{\ac}$ and $\fun_{f_n}\in\mfa_{*,\ac}$. 
Now we show that some of these elements are different from zero: 
Clearly, for any non-zero  $A\in\hmfa_{\cc}$ one can choose such $f\in C_0^{\infty}(\real^s)$ that $A(f)\neq 0$.
Thus we conclude from Lemma~\ref{auxiliary} that $\supp\,\fun(\wt{A(f)}(\ccdot))$ contains $\vep\neq 0$ for some $\fun\in\hmfa_{*,\cc}$. 
Hence,  $\fun(A(f_n))\neq 0$  or, equivalently,  $\fun_{\f_n}(A)\neq 0$.  Now part (b) of the theorem
follows from  Lemma~\ref{auxiliary} and the  inclusions 
\beqa
& &\ov{\underset{\su{A\in\hmfa_{\cc} \\ \fun\in\mfa_*}}\bigcup\supp\,\fun(\wt{A(f_n)}(\ccdot))}\subset \Sp_{\ac}\be,\\
& &\ov{\underset{\su{A\in\mfa \\ \fun\in\hmfa_{*,\cc}}}\bigcup\supp\,\fun_{f_n}(\wt{A}(\ccdot))}\subset \Sp_{\ac}\be^*.
\eeqa

To verify the first statement in part (c), we have to show that for any $A\in\mfa_{\cc}$ the corresponding element
$[A]\in\mfa_{\sic}=\mfa_{\cc}/\mfa_{\ac}$ satisfies $[A](f)=[A(f)]=0$ for any $f\in S(\real^s)$ s.t. 
$\supp\,\tf\cap\{0\}=\emptyset$. To this end, we 
pick a sequence $\{A_m\}_{m\in\nat}$ of elements of $\hmfa_{\cc}$ s.t. $\{A_m(f)\}_{m\in\nat}$ tends to $A(f)$ in norm. From  estimate~(\ref{2regularity}) we obtain
\beq
\int d^sp\,|\fun(\wt{A_m(f)}(\vep))|^2<\infty
\eeq
for any  $\fun\in\hmfa_{*}$. This implies that $A(f)\in\mfa_{\ac}$ i.e. $[A(f)]=0$.

To prove  the second part of (c), one  shows that for any $\fun\in\mfa_{*,\cc}$ the corresponding element
$[\fun]\in\mfa_{*,\sic}$ satisfies $[\fun_f]=0$ for any $f\in S(\real^s)$ s.t. $\supp\,\tf\cap\{0\}=\emptyset$.  
The argument is analogous as above. \qed\\
Part (c) of the above theorem states that  $\Sp_{\sic}\be$ and $\Sp_{\sic}\be^{*}$ are either empty or
consist only  of $\{0\}$. It is an interesting question, whether this latter possibility can be excluded in general.
As a step in this direction, we show in the Appendix, that in theories complying with 
a timelike asymptotic abelianess condition, introduced in \cite{BWa92}, 
\beq
\Sp_{\sic}\be=\Sp_{\sic}\be^*=\emptyset \label{emptiness}
\eeq
for $\hmfa$ slightly smaller than $\lmfa$ chosen here. 
This includes, in particular, the theory of
scalar, non-interacting massive and massless particles\footnote{In the massless case for $s\geq 3$.}. 
In the next section we provide further evidence for triviality of the singular-continuous spectra of $\be$ and $\be^*$:
We  propose a regularity condition, suitable for massive theories,  which implies~(\ref{emptiness}). We also show
that this condition guarantees the existence of particles, if the theory contains a stress-energy tensor.


\section{Structure of the continuous spectrum  and the particle content in QFT}\label{triviality-of-Apc}

In the present section, which is based on Section~2.3 of \cite{Dy08.3}, we  augment the general postulates of quantum field theory, adopted in the previous section, by Conditions~$\B$ and $T$ stated below. The former is a regularity condition, restricting the structure of the continuous spectrum of $\al$ near zero,  while the latter encodes the
presence of a stress-energy tensor among the pointlike-localized fields of the theory. We will show that $\Sp_{\sic}\be=\Sp_{\sic}\be^*=\emptyset$ in theories complying with Condition~$\B$. If, in addition, Condition~$T$ is satisfied, we demonstrate that the theory describes particles in the
sense of non-zero asymptotic functionals.

In order to formulate Condition~$\B$, we have to introduce some terminology:
We define, for any $E\geq 0$ and $C\in\mfa$, the (possibly infinite) quantity 
\beq
\|C\|_{E,1}:=\sup_{\om\in S_E}\int d^sx |\om(\be_{\vx}(C))|, \label{symbol-integrable-semi}
\eeq  
and introduce the following  subspace of $\mfa$
\beq
\CC:=\{\, C\in\mfa \,|\, \underset{E\geq 0}\forall\,\,  \|C\|_{E,1}<\infty  \,\}\label{symbol-integrable-space}
\eeq
which is a natural domain for the asymptotic functionals $\si_{\om}^{(+)}$ mentioned in the Introduction and
defined precisely in (\ref{symbol-asymptotic-functional-appr}) below.
To study the properties of this subspace, we introduce two useful concepts:
First, an operator $B\in\mfa$ 
is called energy-decreasing, if its Arveson spectrum w.r.t. the group of spacetime translation automorphisms
 does not intersect with the closed forward lightcone i.e.
 $\Sp^B\al\,\cap\,\cone=\emptyset$.
Second, an observable $B\in\mfa$ is called almost local, if there exists a net of local operators 
$\{\, B_r\in\mfa(\mco(r))\,|\, r>0\,\}$, s.t. for any $k\in\nat_0$
\beq
\lim_{r\to\infty} r^k\|B-B_r\|=0,
\eeq
where $\mco(r)$ is a double cone of radius $r$, centered at the origin.
After this preparation we state a result, due to Buchholz \cite{Bu90}, which guarantees
non-triviality of $\mfa^{(1)}$ in any local, relativistic quantum field theory.
\bet\cite{Bu90}\label{harmonic} Let $B\in\mfa$ be almost local and energy-decreasing. 
Then, for any  $E\geq 0$, there holds $\|B^*B\|_{E,1}<\infty$.
\eet
\nin Our regularity condition specifies another class of observables from $\CC$.
These operators are of the form $A(g)=\int dt\,g(t) \al_t(A)$, where $A\in\hmfa_{\cc}$ (see definition~(\ref{hmfacc})) and $\tg$ 
is supported in a small neighborhood of zero. More precisely: 
\begin{enumerate}
\item[] \bf Condition $\B$: \rm \label{cond-L1} There exists $\mu>0$ s.t. for any $g\in S(\real)$ with
$\supp\,\tg\subset ]-\mu,\mu[$  and $A\in\hmfa_{\cc}$, 
\begin{enumerate}
\item[(a)] $A(g)\in\mfa^{(1)}$, 
\item[(b)] $\|A(g)\|_{E,1}\leq c_{l,E,r}\|R^l A R^l\|\,\|g\|_1$, for all  $E\geq 0$, $l\geq 0$,
\end{enumerate}
where $R=(1+H)^{-1}$  and  $r>0$ is s.t. $A\in\mfa(\mco(r))$.  
\end{enumerate}
This condition has been verified in massive scalar free field theory in Appendix~D of \cite{Dy08.3}, so it is consistent
with the basic postulates of quantum field theory.  
The quantitative part (b) of this criterion is needed in Theorem~\ref{existence-of-particles} below to prove the existence 
of non-trivial asymptotic functionals in theories admitting a stress-energy tensor. On the other hand, the qualitative part~(a) 
suffices to conclude that the singular-continuous spectrum of the space translation automorphisms is empty. 
\bet\label{triviality}  Let $\real^s\ni\vx\to\be_{\vx}$  be the group of space translation automorphisms acting on the algebra of 
observables  $\mfa$ in a quantum field theory admitting a normal vacuum state $\om_0$ and 
satisfying Condition~$\B$(a). Let $\mfa_*$, $\hmfa_*$ and $\hmfa$ be given by (\ref{spectral-triple}). 
Then $\Spt_{\sic}\be=\Spt_{\sic}\be^*=\emptyset$.
\eet
\proof  Suppose that $A\in\hmfa_{\scc}$. To show that $A\in\mfa_{\ac}$, it suffices to 
verify that for any $E\geq 0$ and $\om\in S_E$
\beq
\int d^sx\,|\om(\be_{\vx}(A))|^2<\infty. \label{weak-ac-quite-new}
\eeq
(This  follows from the Plancherel theorem and  the fact that the distributions $\real^s\ni\vep\to\om(\wt{A}(\vep))$
are compactly supported).  We fix  $E\geq\mu$, where $\mu$ 
appeared in  Condition $\B$, and choose a function
$f\in S(\real)$ s.t. $\tf=(2\pi)^{-\h}$ on $[-E, E]$ and $\supp\,\tf\subset [-2E,2E]$.
With the help of a  smooth partition of unity we can decompose $f$ as follows:
$f=f_-+f_++f_0$, where $\supp\, \tf_-\subset [-2E,-\mu/2]$, $\supp\, \tf_+\subset [\mu/2, 2E]$,
and $\supp\, \tf_0\subset ]-\mu,\mu[$. Then 
\beq
P_EAP_E=P_EA(f)P_E=P_EA(f_-)P_E+P_EA(f_+)P_E+P_EA(f_0)P_E,
\eeq
where the first equality is a consequence of relation~(\ref{continuity-transfer}).
By Condition~$\B$(a),  $A(f_0)$ satisfies the bound~(\ref{weak-ac-quite-new}). 
To the remaining terms we can apply Theorem~\ref{harmonic}, since both $A(f_-)$ and $A(f_+)^*$ are almost local and 
energy-decreasing. This latter fact follows from the equality
\beq
\wt{A(f_-)}(p)=(2\pi)^{\h}\tf_-(p^0)\wt{A}(p^0,\vep)
\eeq
which implies that the support of $\wt{A(f_-)}$ does not intersect with the closed forward lightcone.
(An analogous argument applies to $A(f_+)^*$). We obtain for any 
$\om\in S_E$
\beqa
\int d^sx\, |\om(\be_{\vx}(A(f_-)))|^2
&\leq& \sup_{\om^\prime\in S_E}\int d^sx\, \om^\prime\big(\be_{\vx}(A(f_-)^*A(f_-))\big)\non\\
&=& \|A(f_-)^*A(f_-)\|_{E,1},
\eeqa
where the last expression is finite by Theorem~\ref{harmonic}.
Since an analogous
estimate holds for $A(f_+)$, we conclude that $\hmfa_{\scc}\subset\mfa_{\ac}$ and 
therefore $\mfa_{\scc}=\mfa_{\ac}$ i.e. $\Sp_{\sic}\be=\emptyset$. 
Now suppose that $\fun\in\hmfa_{*,\cc}$. Then, by (\ref{weak-ac-quite-new}), for any $A\in\hmfa$ 
\beq
\int d^sx\,| \be_{\vx}^*\fun(A)|^2<\infty,
\eeq 
which implies that $\fun\in\mfa_{*,\ac}$. We conclude that $\Sp_{\sic}\be^*=\emptyset$.  
\qed

Proceeding to  particle aspects of the theory, we note that the space $\CC$,
equipped with the family of seminorms $\{\,\|\,\cdot\,\|_{E,1}\,|\, E\geq 0\,\}$, 
is a locally convex Hausdorff space and we call the corresponding topology $T^{(1)}$. 
(This is established as in Section 2.2 of \cite{Po04.1}). We define, for any $\om\in S_E$,
a net $\{\si^{(t)}_{\om}\}_{t\in\real_+}$  of functionals on $\CC$ given~by
\beq
\si^{(t)}_{\om}(C):=\int d^sx\, \om(\al_{(t,\vx)}(C)),\quad\quad C\in\CC. \label{symbol-asymptotic-functional-appr}
\eeq
This net satisfies the uniform bound $|\si^{(t)}_{\om}(C)|\leq \|C\|_{E,1}$. Therefore, by the Alaoglu-Bourbaki theorem
(see \cite{Ja}, Section 8.5), it has weak limit points $\si^{(+)}_{\om}$\label{symbol-asymptotic-functional} in the topological 
dual of $(\mfa^{(1)}, T^{(1)})$ which we call the asymptotic functionals. The set of such functionals
\beq
\PC:=\{\, \si^{(+)}_{\om}\, |\, \om\in S_E \textrm{ for some } E\geq 0\, \} \label{symbol-particle-content}
\eeq
will be called the particle content of the theory. This terminology was justified in the Introduction, 
where we argued that the asymptotic functionals should carry information about all the (infra-)particle types
appearing in the theory. A general argument for the existence of non-zero asymptotic functionals has been
given to date only for theories of Wigner particles \cite{AH67}. 
It is now our goal to show that $\PC\neq\{0\}$ not relying  on the Wigner concept of a particle.

Since our argument is based on the existence of a stress-energy tensor, which is postulated
in Condition~$T$ below, we recall the definition and simple properties of pointlike-localized fields:
We set $R=(1+H)^{-1}$\label{symbol-R} and introduce the space of normal functionals with polynomially damped energy 
\beq
\tracei:=\bigcap_{l\geq 0}R^l\trace R^l.\label{symbol-trace-infty}
\eeq
We equip this space with the locally convex topology given by the norms 
$\|\,\cdot\,\|_{l}=\|R^{-l}\,\cdot\, R^{-l}\|$  for  $l\geq 0$. The field content of the theory is 
defined as follows \cite{FH81}
\beq
\Phi_{\FH}:=\{\,\phi\in(\tracei)^* \,|\, R^l\phi R^l\in\bigcap_{r>0} \{R^{l}\mfa(\mco(r))R^{l}\}^{\te{w-cl}}
\textrm{ for some } l\geq 0\,\}, \label{symbol-field-content}
\eeq
where $\te{w-cl}$ denotes the weak closure in $B(\hil)$.
Since the normal vacuum state $\om_0$ is an element of $\tracei$, we can define
\beq
\Phi_{\FH,\scc}:=\{\, \phi\in\Phi_{\FH} \,|\, \om_0(\phi)=0\,\}.
\eeq
There holds the following useful approximation property for the pointlike-localized fields which is due to
Bostelmann \cite{Bo05.1}:
For any $\phi\in\Phi_{\FH,\scc}$ there exists $l\geq 0$ and a net $A_r\in\mfa(\mco(r))$, $r>0$, $\om_0(A_r)=0$, s.t.
\beq
\lim_{r\to 0}\|R^l(A_r-\phi)R^l\|=0. \label{Bostelmann2}
\eeq
Making use of Condition $\B$(b), we also obtain, for any time-smearing function $g\in S(\real)$
s.t. $\supp\,\tg\subset]-\mu,\mu[$,
\beq
\lim_{r\to 0}\|A_r(g)-\phi(g)\|_{E,1}=0. \label{field_approximation}
\eeq
This implies, in particular, that $\|\phi(g)\|_{E,1}<\infty$ for any $\phi\in\Phi_{\FH,\scc}$, which prepares the ground for 
our next assumption:
\begin{enumerate}
\item[] \bf Condition $T$: \rm \label{cond-T} There exists a field $T^{00}\in\Phi_{\FH,\scc}$\label{symbol-T00} which satisfies
\beq
\int d^sx\,\om(\be_{\vx}(T^{00}(g)))=\om(H),\quad\quad \om\in S_E,
\eeq
for any $E\geq 0$ and any time-smearing function $g\in S(\real)$ s.t. $\tg(0)=(2\pi)^{-\fr{1}{2}}$ and 
$\supp\,\tg\subset]-\mu,\mu[$, where $\mu$ appeared in Condition~$\B$.
\end{enumerate}
\nin This condition holds, in particular, in massive scalar free field theory  as shown in Section~B.2 of
\cite{Dy08.3}.  With Conditions~$\B$ and $T$ at hand, we are ready to prove the existence of non-zero
asymptotic functionals. 
\bet\label{existence-of-particles} Suppose that a quantum field theory, admitting a normal vacuum state $\om_0$, 
satisfies Conditions~$\B$ and $T$ and let $\om\in S_E$ be s.t. $\om(H)> 0$. 
Then all the limit points $\si^{(+)}_{\om}$ are non-zero.
\eet
\proof We choose $g\in S(\real)$ as in Condition $T$ and $0<\eps\leq\h|\om(H)|$. Making use of Condition~$\B$(b) and relation~(\ref{field_approximation}), we can find $C\in\CC$  s.t. $\|T^{00}(g)-C\|_{E,1}\leq\eps$.
Then, exploiting Condition~$T$ and invariance of $H$ under time translations, we obtain 
\beq
|\om(H)|=|\int d^sx\,\om\big(\al_{(t,\vx)}(T^{00}(g)) \big)|\leq\eps+|\int d^sx\,\om(\al_{(t,\vx)}(C))|.
\eeq
Thus we arrive at a positive lower bound  $\om(H)\leq 2|\si^{(t)}_{\om}(C)|$   
  which is uniform in $t$. \qed\\
We emphasize that we have proven  more than non-triviality of the particle content - we have verified
that every physical state, with non-zero mean energy, gives rise to a non-trivial asymptotic 
functional. On the other hand, we did not touch upon the problem of convergence of the nets $\{\si^{(t)}_{\om}\}_{t\in\real_+}$
which is essential for their physical interpretation in terms of particle measurements.  
The question, if the energy of the state $\om$ can be reconstructed
from the four-momenta  characterizing the pure particle weights, appearing in the decomposition~(\ref{weltformel1})
of $\si^{(+)}_{\om}$, is another important open problem. Such a result  would be an essential step towards a model-independent understanding of the problem of asymptotic completeness in quantum field theory (cf. discussion in \cite{Bu94}).  Regularity properties of the
continuous spectrum of $\al$ should be of relevance to the study of these issues.    

\section{Conclusions and outlook}\label{section:conclusions}
\setcounter{equation}{0}

In this paper we defined and analyzed the continuous Arveson spectrum
of a group
of isometries $\al$ acting on a  Banach space $\mfa$.  We  introduced 
new notions of the absolutely continuous and singular continuous spectra of $\al$
and  defined the corresponding spectral spaces. By studying  relations between
the spectral concepts on the side of $\al$ and $\al^*$ we found necessary and sufficient
conditions for the pure-point and continuous subspaces to span the entire Banach space.
The sufficient conditions have a natural formulation, if $\mfa$ is a unital $C^*$-algebra
equipped with a distinguished, invariant state $\om_0$. In this setting we established
relations between the continuous spectrum of $\al$ and the spectrum of the implementing group of
unitaries in the GNS representation induced by $\om_0$.    
We verified that in any quantum field theory,
admitting a normal vacuum state,  the group of spacetime translation automorphisms  fits into this algebraic framework. We concluded 
that in a theory of Wigner particles  the singular-continuous spectrum of $\al$  
contains the corresponding mass hyperboloid, while the multiparticle spectrum of the energy-momentum operators
is included in the absolutely continuous spectrum of $\al^*$. Moreover, we found conditions
on the continuous spectrum of $\al$ in a neighborhood of zero which, on the one hand, imply triviality of the
singular-continuous spectrum of the space translation automorphisms, on the other hand entail the existence of 
particles, if the theory contains a stress-energy tensor. 

While this latter assumption is physically reasonable, we feel that the presence of (infra-)particles,
which is a large-scale phenomenon, should not depend on  short-distance properties, like the existence of 
certain pointlike-localized fields. 
It should be possible  to find general necessary and sufficient conditions for non-triviality of the particle content
in  terms of some spectral properties of the group of translation automorphisms. 
In the second step of the analysis these criteria
should be related to physical properties of the theory (e.g. the  phase space structure or the existence of
constants of motion). Since pure particle weights corresponding to different momenta of an infraparticle can 
give rise to inequivalent representations of the algebra of observables, it may be necessary  to look for  
more general spectral concepts than these introduced in the present work. 
Such notions should not depend on the choice of a specific vacuum state,  but rather carry information about  some large class of positive energy representations. First steps in this direction are taken in the Appendix, where
we choose  as $\mfa_*$ the space of energetically accessible functionals (see definition~(\ref{energy-states})), rather than the predual in some vacuum representation. We hope to return to these problems in a future publication.

\bigskip
\bigskip

\noindent{\bf Acknowledgements:}
I would like to thank  Prof.~D.~Buchholz for pointing out to me the absence of
essential spectral concepts in the theory of automorphism groups and the relevance of 
this problem to particle aspects of QFT. I am also grateful to him for numerous
valuable discussions in the course of this work. In particular, for indicating the 
role of the Ergodic Theorem and for pointing out the importance of the singular-continuous spectrum. 

Financial support from 
Deutsche Forschungsgemeinschaft and Graduiertenkolleg 'Mathematische Strukturen in der modernen Quantenphysik'
of the University of G\"ottingen  is gratefully acknowledged. This work  was completed at the TU-M\"unchen, where
it was supported by the DFG grant SP181/25-1.  
Travel grants from Wilhelm und Else Heraeus-Stiftung are also acknowledged as well as a fellowship supported by the Austrian 
Federal Ministry of Science and Research, the High Energy Physics Institute of the Austrian Academy of Sciences and
the Erwin Schr\"odinger International Institute of Mathematical Physics which supported author's
participation in the 4th Vienna Central European Seminar on Particle Physics and Quantum Field Theory.

\newpage

\appendix

\section{\!\!\!\!\!\!\!\!\!\!\,\,\,\,ppendix: Spectral theory of automorphism groups in QFT in the absence of normal vacuum states}\label{Energetically-accessible}
\setcounter{equation}{0}

Theorem~\ref{vacuum-kernel} provides sufficient conditions for the following equalities 
\beqa
\mfa&=&\mfa_{\spp}\oplus\mfa_{\cc}, \label{first-equality}\\  
\mfa_*&=&\mfa_{*,\spp}\oplus\mfa_{*,\cc}\label{second-equality}
\eeqa
which we verified  in Section~\ref{QFT-section} in a large class of examples.
In the first part of this Appendix we describe situations where decompositions~(\ref{first-equality}), (\ref{second-equality}) fail. 
In particular, we show that in the case of the group of space translation automorphisms $\be$, acting on the algebra of observables in quantum field theory,
equality~(\ref{first-equality}) fails, if  $\mfa_*$ contains more than one vacuum state.
This occurs in some low-dimensional massless theories, if $\mfa_*$ is chosen as the space of energetically accessible functionals,  defined in~(\ref{energy-states}) below. On the other hand, for theories complying with Condition~$\Ab$, stated below,  equalities~(\ref{first-equality}),   
and (\ref{second-equality}) hold for this choice of $\mfa_*$. In the second part of the Appendix, which relies on results from \cite{BWa92}, we analyze briefly the continuous spectra of $\be$ and $\be^*$. It turns out that their singular-continuous parts are empty in this setting.  

We start from the following simple observation: 
\bep\label{general-vacuum} Let  $\real^s\ni\vx\to\be_{\vx}$ be the group of space translation automorphisms acting on the algebra of 
observables  $\mfa$ in a quantum field theory (possibly without normal vacuum states). Then  $\mfa_{\spp}=\Spant\,\{I\}$
for any $\mfa_*$ satisfying Condition~$\C$. 
\eep
\proof Suppose that $A\in\mfa$ is an eigenvector i.e.
\beq
\be_{\vx}(A)=e^{-i\veq\vx}A, \quad \vx\in\real^{s}
\eeq
for some $\veq\in\real^{s}$. Then $A$ belongs to the center of $\mfa$, since locality gives
\beq
\|[A,B]\|=\lim_{|\vx|\to\infty}\|[\al_{\vx}(A),B]\|=0,\quad\quad B\in\mfa.
\eeq
The irreducibility assumption ensures that the center of $\mfa$ consists only of
multiples of the identity. \qed\\ 
In view of Theorem~\ref{main-ppc-theorem}~(a), equality~(\ref{second-equality})  fails, in particular, 
if  $\dim\, \tmfa_*(\{0\})<\dim\,\tmfa(\{0\})$. Let us consider the group of space translation automorphisms $\be$ acting on the algebra of observables
$\mfa$ in a quantum field theoretic model which does not admit normal vacuum states, (see  \cite{BHS63}
for an example). Thus, choosing $\mfa_*=\trace$, we obtain $\dim\,\tmfa_*(\{0\})=0$, whereas  Proposition~\ref{general-vacuum} 
gives $\dim\,\tmfa(\{0\})=1$.

On the other hand, equality~(\ref{first-equality}) fails when $\dim\, \tmfa_*(\{0\})>\dim\,\tmfa(\{0\})$. To exhibit an example,
let us choose as $\mfa_*$ the space of energetically accessible functionals
\beq
\traca:=\Span\big\{\bigcup_{E\geq 0} S_E^{\te{w$^*$-}\cl}\big\}^{\no\cl}, \label{energy-states}
\eeq
where $\te{w$^*$-}\cl$ denotes the closure in the weak$^*$ topology of $\mfa^*$. (Clearly, this space  
satisfies Condition~$\C$). It is well known, that massless free field theory in $s=2$ dimensional space has an infinite family
of vacuum states in $\traca$ \cite{BWa92}. Hence, by Theorem~\ref{general-vacuum}, $\mfa\neq\mfa_{\spp}\oplus\mfa_{\cc}$
in this situation.

However, there exists a large class of theories, in which the choice $\mfa_*=\traca$ entails 
equalities~(\ref{first-equality}) and (\ref{second-equality}). These are, in particular, models which satisfy the
following asymptotic abelianess assumption, proposed by Buchholz and Wanzenberg \cite{BWa92}.  
\begin{enumerate}
\item[] \bf Condition $\Ab$: \rm  There exists a norm dense subspace $\D\subset\lmfa$ s.t. for any
$A\in \D$ there exists some positive number $1\leq r<s$, s.t. for all $\Phi\in\hil$
\beq
\sup_{x_0}\int d^sx\|[A^*,\al_{x_0,\vx}(A)]\Phi\|^r<\infty.
\eeq 
\end{enumerate}
These authors have shown that Condition~$\Ab$ holds in massive (for $s\geq 1$) and massless (for $s\geq 3$) free field theory. 
Moreover, it was verified in \cite{BWa92} that in theories complying with Condition~$\Ab$ there exists a distinguished vacuum 
state $\om_0$ in $\traca$ s.t. for $A\in\mfa$
\beq
\te{w$^*$-}\lim_{|\vx|\to\infty}\be_{\vx}(A)=\om_0(A)I.
\eeq
(Other conditions which imply this property can be found in \cite{Dy08.1, Dy09}).
By a slight modification of the discussion from Section~4 of \cite{BWa92},
we obtain that for any  $A\in\ker\,\om_0$, $\fun\in\traca$ and $\veq\in\real^s$ 
\beq 
\underset{K\nearrow \real^{s}}\lim
\fr{1}{|K|}\int_{K} e^{i\veq\vx}\fun(\be_{\vx}(A))\,d^{s}x=0. \label{Ergodicity}
\eeq
Hence $\ker\,\om_0\subset\mfa_{\cc}$ and, consequently, $\om_0$ is the unique element of $\traca$ invariant 
under the action of $\be^*$. Thus the decomposition of $\mfa$ and $\mfa_{*}$ into the pure-point and
continuous subspaces is given by Theorem~\ref{vacuum-kernel}. We summarize:
\bet\label{key-assumption-verification-new} Let $\real^{s}\ni \vx\to\be_{\vx}$ be the group of space translation 
automorphisms acting on the algebra of observables $\mfa$ in a quantum field theory satisfying 
 Condition~$\Ab$. Let $\mfa_*=\traca$ and let $\om_0$ be the unique vacuum state in $\traca$.
Then $\ker\om_0\subset\mfa_{\cc}$ and, consequently,
\begin{enumerate} 
\item[(a)] $\mfa=\mfa_{\spp}\oplus\mfa_{\cc}$, where $\mfa_{\spp}=\Spant\,\{I\}$,   $\mfa_{\cc}=\ker\,\om_0$,
\item[(b)] $\mfa_*=\mfa_{*,\spp}\oplus\mfa_{*,\cc}$, where $\mfa_{*,\spp}=\Spant\,\{\om_0\}$, $\mfa_{*,\cc}=\ker\, I$.
\end{enumerate}
\eet
The analysis of the absolutely continuous and singular-continuous spectrum is performed similarly as
in Theorem~\ref{refined-theorem-old}. However, the norm dense subspaces $\hmfa\subset\mfa$ and $\hmfa_*\subset\mfa_*$,
are now chosen as follows
\beqa
& &\hmfa=\D, \label{hmfaa} \\
& &\hmfa_*=B(\hil)_{*,\te{bd}}^{(a)}:=\Span\big\{\bigcup_{E\geq 0} S_E^{\te{w$^*$-}\cl}\big\},\label{hmfaas}
\eeqa
where $\D$ appeared in Condition~$\Ab$. In the present case we are able to show that the 
singular-continuous spectra of $\be$ and $\be^*$ are empty.  
\bet\label{refined-theorem-old-old} Let $\real^{s}\ni \vx\to\be_{\vx}$ be the group of space translation 
automorphisms acting on the algebra of observables $\mfa$ in a quantum field theory satisfying 
 Condition~$\Ab$. Let $\mfa_*$, $\hmfa$ and $\hmfa_*$ be given by (\ref{energy-states}), (\ref{hmfaa}) and (\ref{hmfaas}), respectively. 
  Then there holds:
\begin{enumerate}
\item[(a)] $\Spt_{\spp}\be=\Spt_{\spp}\be^{*}=\{0\}$,  
\item[(b)] $\Spt_{\ac}\be=\Spt_{\ac}\be^*=\real^s$,
\item[(c)] $\Spt_{\sic}\be=\Spt_{\sic}\be^*=\emptyset$.   
\end{enumerate}
\eet
\proof Part (a) follows from Theorem~\ref{key-assumption-verification-new}. 
To prove statements (b) and (c), we define the subspaces 
\beqa
\hmfa_{\cc}^{(a)}&:=&\{\,A\in\hmfa \,|\, \om_0(A)=0\}, \label{dense-subspace-one-more}\\
\hmfa_{*,\cc}^{(a)}&:=&\{\, \fun\in\hmfa_{*} \,|\, \fun(I)=0\} \label{dense-subspace-two-more}
\eeqa
which are norm dense in $\mfa_{\cc}$ and $\mfa_{*,\cc}$, respectively. We choose $A\in\hmfa_{\cc}^{(a)}$, 
$f\in S(\real^s)$ s.t. $\tf$ vanishes in some neighborhood of zero and set $A(f):=\int d^sx\,f(\vx)\be_{\vx}(A)$. 
Then we obtain  from Lemma~2.1 of \cite{BWa92} the bound   
\beq
\|P_EA(f)P_E\|\leq c\big(\int d^sp\, |\vep|^{-(s-\eps)}|\tf(\vep)|^2 \big)^{1/2}
\eeq 
for some constants $c>0$, $\eps >0$ independent of $f$. Noting that for any $\om\in S_E^{\we\cl}$,  
$|\om(A(f))|\leq\|P_EA(f)P_E\|$ holds, making use of the assumption that $\be$ acts strongly continuously
on $\mfa$ to exchange the action of the state $\om$ with integration and proceeding as in \cite{BWa92}, p.581,
we obtain 
\beq
\om(\be_{\vx}(A))=l(\vx)+\om^\prime_0(A),
\eeq
where $\tilde l\in L^1(\real^s,d^sp)$ and $\om^\prime_0=\te{w$^*$-}\lim_{|\vx|\to\infty}\be^*_{\vx}\om$ is an element of $S_E^{\we\cl}$,
invariant under the action of $\be^*$. So, by Theorem~\ref{key-assumption-verification-new}~(b), $\om_0^\prime=\om_0$ and consequently
\beq
\underset{\su{A\in\hmfa_{\cc}^{(a)} \\ \fun\in\hmfa_{*,\cc}^{(a)}  }}\forall  \int d^sp\, |\fun(\wt{A}(\vep))|<\infty.
\eeq 
We conclude, that $\mfa_{\cc}=\mfa_{\ac}$ and $\mfa_{*,\cc}=\mfa_{*,\ac}$, which proves (c). 
Part (b) follows from the inclusions
\beqa
& &\ov{\underset{\su{A\in\hmfa_{\cc}^{(a)} \\ \fun\in\mfa_*}}\bigcup\supp\,\fun(\wt{A}(\ccdot))}\subset \Sp_{\ac}\be,\\
& &\ov{\underset{\su{A\in\mfa \\ \fun\in\hmfa_{*,\cc}^{(a)} }}\bigcup\supp\,\fun(\wt{A}(\ccdot))}\subset \Sp_{\ac}\be^*
\eeqa
and from Lemma~\ref{auxiliary}. To apply this latter fact, we note that $\hmfa_{\cc}^{(a)}\subset\hmfa_{\cc}$ and  $\hmfa_{*,\cc}^{(a)}\supset\hmfa_{*,\cc}$, 
where $\hmfa_{\cc}$ and $\hmfa_{*,\cc}$  are given by (\ref{hmfacc}) and (\ref{hmfascc}), respectively.
\qed\\
The above result has the following immediate corollary: After adding Condition~$\Ab$ to the assumptions of Theorem~\ref{refined-theorem-old}
and choosing $\hmfa=\D$,  part (c) of this theorem can be strengthened to $\Sp_{\sic}\be=\Sp_{\sic}\be^*=\emptyset$.


\end{document}